\documentclass[a4paper,twocolumn,english,aps,prl,superscriptaddress]{revtex4-1}
\usepackage[T1]{fontenc}
\usepackage[latin9]{inputenc}
\usepackage{color}
\usepackage{babel}
\usepackage{bm}
\usepackage{amsmath}
\usepackage{amssymb}
\usepackage{graphicx}
\usepackage[unicode=true,pdfusetitle,
 bookmarks=true,bookmarksnumbered=false,bookmarksopen=false,
 breaklinks=false,pdfborder={0 0 1},backref=false,colorlinks=false]
 {hyperref}
\usepackage{breakurl}

\makeatletter

\usepackage{babel}
\usepackage{babel}

\makeatother
\begin{document}
\title{Transition from metal to higher-order topological insulator driven by random flux}
\author{Chang-An Li}
\email{changan.li@uni-wuerzburg.de}

\affiliation{Institute for Theoretical Physics and Astrophysics, University of
W\"urzburg, 97074 W\"urzburg, Germany}
\author{Song-Bo Zhang}
\email{songbo.zhang@physik.uzh.ch}

\affiliation{Institute for Theoretical Physics and Astrophysics, University of
W\"urzburg, 97074 W\"urzburg, Germany}
\author{Jan Carl Budich}
\affiliation{Institute of Theoretical Physics, Technische Universit\"at Dresden, 01062 Dresden, Germany}
\affiliation{W\"urzburg-Dresden Cluster of Excellence ct.qmat, Germany}

\author{Bj\"orn Trauzettel}
\affiliation{Institute for Theoretical Physics and Astrophysics, University of
W\"urzburg, 97074 W\"urzburg, Germany}
\affiliation{W\"urzburg-Dresden Cluster of Excellence ct.qmat, Germany}
\date{\today}
\begin{abstract}
Random flux is commonly believed to be incapable of driving full metal-insulator transitions in non-interacting systems.
Here we show that random flux can after all induce a full metal-band insulator
transition in the two-dimensional Su-Schrieffer-Heeger model. Remarkably, we find that the resulting insulator can be an extrinsic higher-order
topological insulator with zero-energy corner modes in proper regimes, rather than a conventional Anderson insulator.
Employing both level statistics and finite-size scaling analysis, we characterize the metal-band insulator transition and numerically extract its critical exponent as $\nu=2.48\pm0.08$.
To reveal the physical mechanism underlying
the transition, we present an effective band structure picture based on the random flux averaged Green's function.
\end{abstract}
\maketitle
\textit{Introduction.}---Disorder, being present in
most physical systems, constitutes a broad field of physics research.
As one of its most salient effects, random potential disorder can induce metal-Anderson insulator transitions
in various systems~\cite{Anderson58pr,Evers08rmp,Sanchez-Palencia10NP,Schwartz07nature,Chabe08prl}, prominently topological phase transitions~\cite{LiJian09prl,Groth09prl,JiangH09prb,GuoHM10prl}, as recently observed in cold-atom and photonic systems
\cite{Meier19Science,Stutzer18nature}. Random flux is another generic
type of disorder that has been widely investigated in two-dimensional
(2D) electron systems~\cite{Cerovski01prb,Furusaki99prl,Aronov94prb,Taras00prl,Sheng95prl,LiuDZ95prb, ZhangSC94prl, Gade93npb,Sugiyama93prl,Avishai93prb,PALee81prl,AnJ01prb,Foster08prb}.
Yet, it is believed that random flux is unable to drive
a system with chiral symmetry from metal to Anderson insulator if the Fermi energy locates precisely at zero: instead it localizes all states except the ones
at the band centre~\cite{Cerovski01prb,Furusaki99prl}.
Moreover, the interplay between random flux and topology has barely
been explored.

\begin{figure}
\includegraphics[width=1\linewidth]{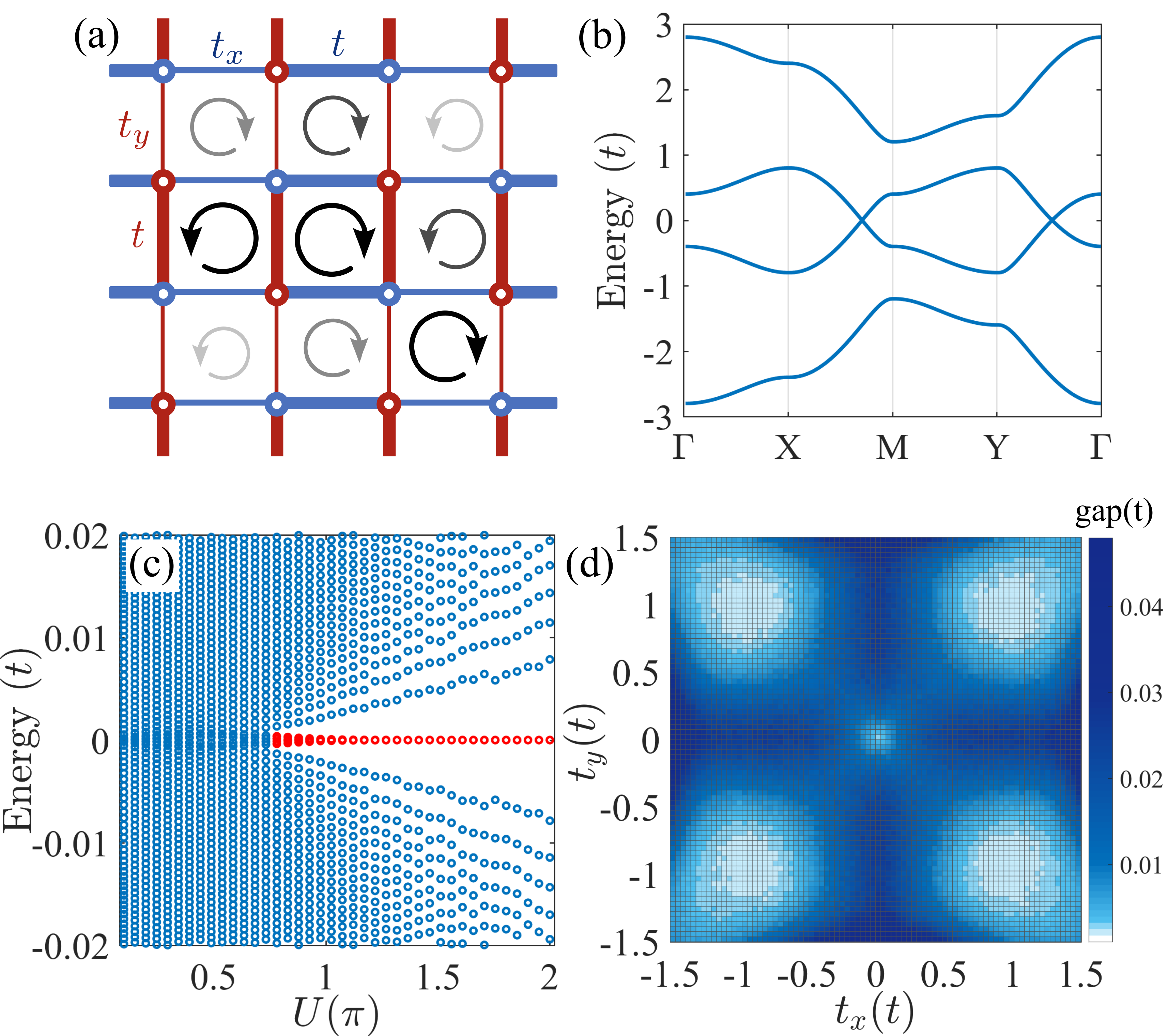}
\caption{(a) Schematic of the 2D SSH model with random flux. Blue(red)
thick and thin bonds mark dimerized hopping amplitudes in $x(y)$-direction. The round arrow (with different sizes and opacity) in each plaquette indicates the random flux. (b) Energy spectrum of the model without random flux for  $(t_{x},t_y)=(0.2t,0.6t)$. (c) Disorder-averaged spectrum as a function of $U$ for $(t_{x},t_y)=(0.2t,0.6t)$.  For large $U$, the system acquires a bulk gap that protects four zero-energy modes (red). (d) Density plot of the directly disorder-averaged gap as a function of $t_{x}$ and $t_{y}$ at $U=2\pi$.  The dimension of the system is $L\equiv L_{x}=L_{y}=30$ with open (periodic) boundaries in (c) {[}(d){]}. Here, 200 random-flux configurations are considered. \label{fig:lattice}}
\end{figure}

In this work, we discover a random-flux driven
metal-band insulator transition. To this end, we add random flux to the 2D Su-Schrieffer-Heeger (SSH) lattice model [Fig.\ \ref{fig:lattice}(a)] which has attracted broad interest recently \cite{LiuF17prl,Benalcazar17Science,BBH17prb}. In the absence of random flux, this model has been realized in different physical platforms \cite{Schindler18NP,Imhof18np,Serra-Garcia18nature,XieBY19prl,QiY20prl,Peterson18nature,Ni19nm,ChenXD19prl}, and sparked the rapidly developing field of higher-order topological phases \cite{Langbehn17prl,SongZD17prl,Schindler18SA,Ezawa18prl,Geier18prb,Trifunovic19prx, Imhof18np,Serra-Garcia18nature,XieBY19prl,QiY20prl,Peterson18nature,Ni19nm,ChenXD19prl, LiuF19prl,XWLuo19prl,Ghorashi20prl,WangHX20prl,Kudo19prl,Volpez19prl,WangZJ19prl,YanZB18prl, ZhangRX20prl,LiCA20prl,ChenR20prl,YangYB21prb,SBZhang20PRR,LiT21arxiv}.
Importantly, the existence of a metallic phase in the clean 2D SSH model and its rich topological properties due to non-trivial inner  degrees of freedom provide a promising playground for revisiting the issue of random-flux driven transitions in the context of topological band structures.

Remarkably, we find that the spectrum of the
system acquires a finite bulk gap in a broad parameter range when exceeding a critical strength of random flux [Figs.~\ref{fig:lattice}(c) and \ref{fig:lattice}(d)], thus transforming from a metallic phase to a  band insulator. This metal-band insulator transition
is confirmed and carefully analyzed by employing energy level statistics
and finite-size scaling theory. The corresponding critical exponent
is estimated to be $\nu=2.48\pm0.08$. Interestingly, we find that the
band insulator induced by random flux can be an extrinsic higher-order
topological insulator (HOTI) by calculating the  topological index $q_{xy}$
and identifying the corresponding boundary signatures. Furthermore,
with an effective band structure picture based on the flux-averaged Green's function, we show that the metal-band insulator transition can be attributed to the emergence of strongly momentum-dependent flux-induced terms that have a non-trivial matrix structure in the effective Hamiltonian. By contrast, such an interplay of random flux and internal degrees of freedom in the unit cell is absent in the conventional random-flux model.

\textit{2D SSH lattice with random flux.}---As
visualized in Fig.~\ref{fig:lattice}(a), the 2D SSH lattice model
features dimerized hopping amplitudes along both $x$- and $y$-directions~\cite{LiuF17prl}. In the absence of disorder, it can be described by the Hamiltonian
\begin{alignat}{1}
H_{0}({\bf k})= & (t_{x}+t\cos k_{x})\tau_{1}\sigma_{0}-t\sin k_{x}\tau_{2}\sigma_{3}\nonumber \\
 & +(t_{y}+t\cos k_{y})\tau_{1}\sigma_{1}-t\sin k_{y}\tau_{1}\sigma_{2},\label{eq:Hamiltonian}
\end{alignat}
where $\tau$ and $\sigma$ are Pauli matrices for different degrees
of freedom within a unit cell; ${\bf k}=(k_{x},k_{y})$ is the 2D wave-vector; $t$ and $t_{x}$$(t_{y}$) denote the two staggered
hopping strengths in $x$$(y)$-direction. For simplicity, we put
the lattice constant to unity and assume $t>0$. Note
that $k_{x}$ and $k_{y}$ are decoupled in Eq.~\eqref{eq:Hamiltonian}.
The total Hamiltonian can be recast as the sum of
two SSH models along $x$- and $y$-directions, respectively, i.e.,
$H_{0}(\mathbf{k})=H_{x}(k_{x})+H_{y}(k_{y})$. The matrices $\tau_{1}\sigma_{0}$
and $\tau_{2}\sigma_{3}$ contained in $H_{x}(k_{x})$ anticommute
with each other. The same holds for the matrices $\tau_{1}\sigma_{1}$
and $\tau_{1}\sigma_{2}$  contained in $H_{y}(k_{y})$. However, the two blocks commute with each other, i.e., $[H_{x}(k_{x}),H_{y}(k_{y})]=0$. As a consequence, the four energy bands of Eq.~\eqref{eq:Hamiltonian} are given by
$E_{\eta}^{\pm}=\pm[\epsilon_{x}(k_{x})+(-1)^{\eta}\epsilon_{y}(k_{y})]$
with $\epsilon_{\alpha}(k_{\alpha})=\sqrt{t_{\alpha}^{2}+2t_{\alpha}t\cos k_{\alpha}+t^{2}}$, $\alpha\in\{x,y\}$ and $\eta\in\{1,2\}$. When $||t_{x}|-|t_{y}||<2t$, the system is in a metallic phase at low energies [Fig.\ \ref{fig:lattice}(b)].
The model has $C_{2v}$ group symmetry in general ($t_{x}\neq t_{y}$).
Moreover, it respects chiral symmetry $\gamma_{5}H_{0}({\bf k})\gamma_{5}^{-1}=-H_{0}({\bf k})$
with the chiral operator $\gamma_{5}=\tau_{3}\sigma_{0}$.
In the clean case, the constituting 1D blocks along $x$-
and $y$-directions are topologically nontrivial when $|t_{x}|<t$ and $|t_{y}|<t$, respectively.
This property can be identified by symmetry indicators based on the
symmetry representations at high-symmetry points in Brillouin
zone that are described in Refs.~\cite{Benalcazar19prb,Krutoff17prx,Po17nc,Li2021SM}. We note that there
may be corner-localized bound states in the bulk continuum, while their stability needs to be protected by $C_{4v}$ symmetry~\cite{Benalcazar20prb,Cerjan21prl} which corresponds to
$t_{x}=t_{y}$ in Eq.\ \eqref{eq:Hamiltonian}.

We now add random flux to the model such that each plaquette encloses
a flux that has random values drawn from a uniformly distributed interval
$[-U/2,U/2]$, as illustrated in Fig.\ \ref{fig:lattice}(a). Here,
$U$ is the strength of random flux within the range of $[0,2\pi]$,
in units of the magnetic flux quantum $\Phi_{0}=hc/e$ \cite{Note1}. The random
flux generates random Peierls phases in the hopping matrix elements.
Thus, time reversal symmetry is broken. However, chiral symmetry is
still preserved and plays a crucial role in the metal-insulator transition
as we elaborate below. Note that when each plaquette encloses a $\pi$ flux uniformly, the system is deformed to the Benalcazar-Bernevig-Hughes (BBH) model \cite{Benalcazar17Science,BBH17prb}

\textit{Metal-band insulator transition driven by random
flux.}---Next, we demonstrate the existence of random-flux driven metal-band insulator transitions
in the 2D SSH model by employing level statistics~\cite{Wigner51,Dyson62jmp}.
In the presence of chiral symmetry, the model falls into
the chiral unitary universality class, i.e., AIII in AZ classification~\cite{Altland97}. The insulating and metallic phases can be distinguished by inverse participation ratio (IPR) \cite{LiX17prb,RoyS21prl,Padhan22prb} and level spacing ratio (LSR) \cite{YangYB21prb,Oganesyan07prb}. The IPR is defined by the eigenstates $\phi_{n}({\bf R},\zeta)$ of the system as
\begin{alignat}{1}
I_{n}= & \sum_{{\bf R}}\sum_{\zeta=1}^{4}|\phi_{n}({\bf R},\zeta)|^{4},
\end{alignat}
where the sums run over all unit
cells labeled by ${\bf R}$ and the inner degrees of freedom $\zeta$
within a unit cell. The subscript $n$ stands for the $n$-th state with
the corresponding eigenenergy $E_{n}$ listed in ascending order.
The LSR is defined in terms of the spectrum as\ \cite{Oganesyan07prb}
\begin{equation}
r_{n}=\frac{\mathrm{min}(s_{n},s_{n-1})}{\mathrm{max}(s_{n},s_{n-1})},
\end{equation}
where $s_{n}\equiv E_{n+1}-E_{n}$ is the difference between two adjacent
energy levels. Both the averaged IPR $\langle I\rangle$ and LSR $\langle r\rangle$ take different
values in the metallic and insulating limits, thus providing important
tools to characterize metal-insulator transitions.

\begin{figure}
\includegraphics[width=1\linewidth]{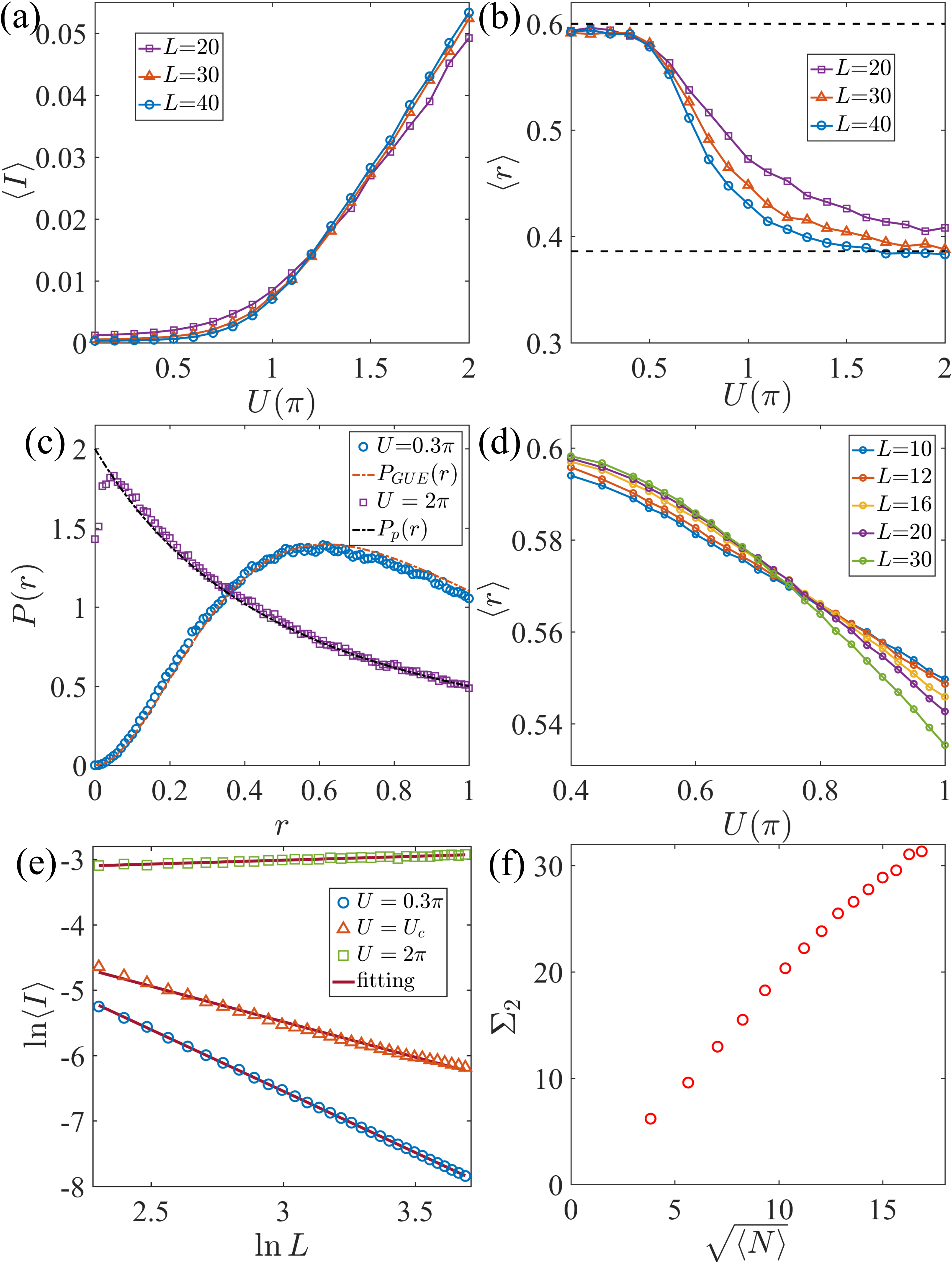}
\caption{(a) Averaged IPR $\langle I\rangle$ as a function of $U$ for $L=20$,
$30$, and $40$, respectively. (b) Averaged LSR $\langle r\rangle$
as a function of $U$ for $L=20$, $30$, and
$40$, respectively. (c) Distribution of LSR $P(r)$ in different
limits. (d) Averaged LSR $\langle r\rangle$ near the critical point
as a function of $U$ for various $L$. (e) Scaling behavior of IPR
in the metallic (blue) and insulating (green) phases, and at the critical
point $U_{c}\approx0.75\pi$ (red). (f) Number variance $\Sigma_{2}$ as a function
of $\sqrt{\langle N\rangle}$ at the critical point.
We consider $N_{E}=16$ and 4000 random-flux configurations in (a,b,c,e). The parameters $t_{x}=0.2t$, $t_{y}=0.6t$ and periodic boundary conditions are chosen for all plots.  \label{fig:scaling}}
\end{figure}

We show below that the level statistics smoothly cross over between
the two limits as the random flux drives the system from a metallic to an insulating phase. Due to the presence of chiral symmetry, the eigenenergies
of the system come in pairs ($\pm E_{n}$). For illustration, we take
$t_{x}=0.2t$ and $t_{y}=0.6t$ and consider an energy window containing
$N_{E}$ energy levels around $E=0$. Figure\ \ref{fig:scaling}(a)
displays $\langle I\rangle$ as a function of $U$. Clearly, $\langle I\rangle$ increases monotonically from nearly
zero in the small $U$ limit to finite values for large $U.$ This
indicates that the system transits from a metallic (with vanishing
$\langle I\rangle$) to an insulating phase (with finite $\langle I\rangle$).
Concomitant with the transformation of $\langle I\rangle$, we also
observe that $\langle r\rangle$ decreases smoothly from a universal
value $0.6$ at small $U$ ($\simeq0$) to another universal value
$0.386$ at large $U$ ($\simeq2\pi$), as shown in Fig.~\ref{fig:scaling}(b).
For sufficiently large $L$, the numerical values approach the universal
constants in both limits of $U$. These results agree with those obtained for the uncorrelated Poisson ensemble in the insulating phase ($\langle r\rangle_{\mathrm{ins}}\approx0.386$)~\cite{Oganesyan07prb} and the unitary ensemble in the metallic phase
($\langle r\rangle_{\mathrm{met}}\approx0.6$)~\cite{Atas13prl},
respectively.

%This behavior strongly indicates a metal-insulator transition driven by random flux.

To better illustrate the transition, we analyze the
probability distribution $P(r)$ of LSR~\cite{Note-r-coninutity}.
As shown in Fig.\ \ref{fig:scaling}(c), $P(r)$ also exhibits universal
but different forms in the small and large $U$ regimes, respectively~\cite{Note2}. For small $U$, we find that $P(r)$ can be well described
by the distribution function of the Gaussian unitary ensemble (GUE)
$P_{\text{GUE}}(r)=\frac{81\sqrt{3}}{2\pi}\frac{(r+r^{2})^{2}}{(1+r+r^{2})^{4}}$~\cite{Atas13prl}. This finding supports that the system is in a metallic
phase. For large $U$, $P(r)$ instead resembles the uncorrelated
Poisson distribution $P_{\text{p}}(r)=\frac{2}{(1+r)^{2}}$, which
again hallmarks an insulating phase~\cite{Oganesyan07prb}. These results
provide direct evidence that the system undergoes a metal-insulator
transition by increasing $U$. This metal-insulator transition
is generic for parameters fulling $||t_{x}|-|t_{y}||<2t$, $t_{x}\neq t$, and $t_{x}\neq t$
\cite{Note3}. It is, however, absent for $t_{x}=t_{y}=t$ which corresponds to
the conventional random-flux model~\cite{Li2021SM}.
We note that the band gap opening  by random flux [see Fig.~\ref{fig:lattice}(c)] may modify the statistical behavior of low energy states close to the band center.

\textit{Critical exponent.}---Critical exponents
are keys  for characterizing continuous phase transitions.
To identify the critical exponent $\nu$ and critical random-flux strength $U_{c}$, we perform a finite-size scaling analysis of the averaged LSR $\langle r\rangle$~\cite{Laumann14prl,Luitz15prb,LuoX21prl}. According to the single-parameter
scaling theory, $\langle r\rangle$ shows a size-independent value
at $U=U_{c}$. Concentrating around the zero energy, we fix
the energy window to capture $10\%$ of the eigenvalues and choose
the number of random-flux configurations in such a way  that the total
eigenvalue number reaches $5\times10^{7}$. As shown
in Fig.\ \ref{fig:scaling}(d), $\langle r\rangle$ increases as the system size
$L$ grows before the transition whereas it decreases as $L$
grows after the transition. The scaling argument near $U_{c}$ states
that $\langle r\rangle$ can be described by a universal function
of the form $F(f_{1}(u)L^{1/\nu},f_{2}(u)L^{-y})$ characterized by
$\nu$, where $u\equiv(U-U_{c})/U_{c},$ and $y$ is an auxiliary exponent;
$f_{1}(u)L^{1/\nu}$ and $f_{2}(u)L^{-y}$ stand for relevant and
irrelevant length-scale corrections, respectively \cite{Slevin14njp,LuoX21prl}. Close to $U_{c}$,
we expand $f_{\eta}(u)=\sum_{j=0}^{m_{\eta}}a_{j}^{\eta}u^{j}$ with
$\eta\in\{1,2\}$. Thus, $\nu$ and $U_{c}$ can be identified by
fitting the Taylor expansion of the function $F$ near the critical
point \cite{Slevin14njp,LuoX21prl}. Thereby, we identify the critical
exponent of the random-flux driven metal-band insulator transition as $\nu=2.48\pm0.08$.
This critical exponent is close to that of integer
quantum Hall transitions with $\nu\approx2.59$ \cite{Slevin09prb}.
In contrast to $\nu$ \cite{Note_exponent}, the critical strength $U_{c}$ depends explicitly
on the parameters $t_{x}$ and $t_{y}$. For the parameters considered
in Fig.\ \ref{fig:scaling}(d), we find $U_{c}\approx0.75\pi$, in
accordance with the gap opening [Fig.\ \ref{fig:lattice}(c)].

\textit{Fractal dimension and spectral rigidity.}---At
the critical point, the wavefunctions of the system show multifractality
due to strong fluctuations~\cite{Evers00prl,Mirlin00prb}. The multifractality
gives rise to one of the fractal dimensions $d_{2}$
defined through the scaling behavior $\langle I\rangle\propto L^{-d_{2}}$.
Figure~\ref{fig:scaling}(e) displays $\ln\langle I\rangle$ as a function of $\ln L$ at small,
large, and critical values of $U$, respectively. At the critical point $U=U_{c}$
(triangles), we can extract $d_{2}=1.085\pm0.034$. At $U=0.3\pi$
(circles) and $U=2\pi$ (squares), we obtain $d_{2}=1.880\pm0.006$
and $0.119\pm0.008$, which are close to the  values of an  ideal metal (corresponding to $d_{2}=2$)
and an insulator (corresponding to $d_{2}=0$), respectively.

The spectral rigidity is also related to the wavefunction multifractality.
It is defined as the level number variance $\Sigma_{2}\equiv\langle N^{2}\rangle-\langle N\rangle^{2}$
in an energy window, where $\langle N\rangle$ is the disorder-averaged
number of energy levels within this window. For conventional Anderson
transitions, $\Sigma_{2}\propto\langle N\rangle$ at the critical
point when the energy window is sufficiently large. The ratio $\chi\equiv\Sigma_{2}/\langle N\rangle$ defines the compressibility of the spectrum. It is conjectured that
$d_{2}$ is related to $\chi$ by the relation $\chi=(2-d_{2})/4$
in 2D~\cite{Chalker96jetp,Chalker96prl}. However, our scaling law follows instead $\Sigma_{2}\propto\sqrt{\langle N\rangle}$
[Fig.~\ref{fig:scaling}(f)], resembling the complex Ginibre
ensemble~\cite{Ginibre00jmp}. This behavior may be due to the fact
that the random flux gives a complex matrix ensemble. Thus, $\chi$ goes to zero in the large $N$ limit, and the aforementioned
conjecture fails in our system.

\textit{Effective band structure picture for the
metal-band insulator transition.}---%Now, we show that the random flux-driven insulating phase is indeed a band insulator. 
To reveal the underlying mechanism,
we average the Green's function over many random-flux configurations, so as to effectively restore lattice translation-invariance
and derive the self-energy $\Sigma({\bf k})$ due to the
random flux \cite{Li2021SM,Note6}. We find that $\Sigma({\bf k})$ not only modifies
the coefficient functions of the matrices in the original Hamiltonian
{[}\textit{c.f.} Eq.~\eqref{eq:Hamiltonian}{]} but also introduces additional
terms associated with new matrices $\tau_{1}\sigma_{3}$ and $\tau_{2}\sigma_{0}$
(that also appear in the BBH model). This feature can be understood
in terms of higher-order scattering processes induced by random
flux. It is intimately related to the interplay between the internal  degrees of freedom of the model and the random flux that couples directly
to momentum in the system. Consequently, $\Sigma({\bf k})$ decisively
depends on momentum. These observations indicate that the effective
Hamiltonian $H_{\mathrm{eff}}({\bf k})\equiv H_{0}({\bf k})+\Sigma({\bf k})$
for the system with random flux can be regarded as a mixture of the
2D SSH and BBH models. Remarkably, a band gap for strong
$U$ can be directly revealed by the effective band structure of $H_{\mathrm{eff}}({\bf k})$~\cite{Li2021SM}. The critical value  $U_c$ of random flux strength  obtained here is consistent with the numerical result in Fig.\ \ref{fig:lattice}(c).  In this sense, the random flux generates a band insulator by opening an effective gap in the bulk after the transition.

\begin{figure}
\includegraphics[width=1\linewidth]{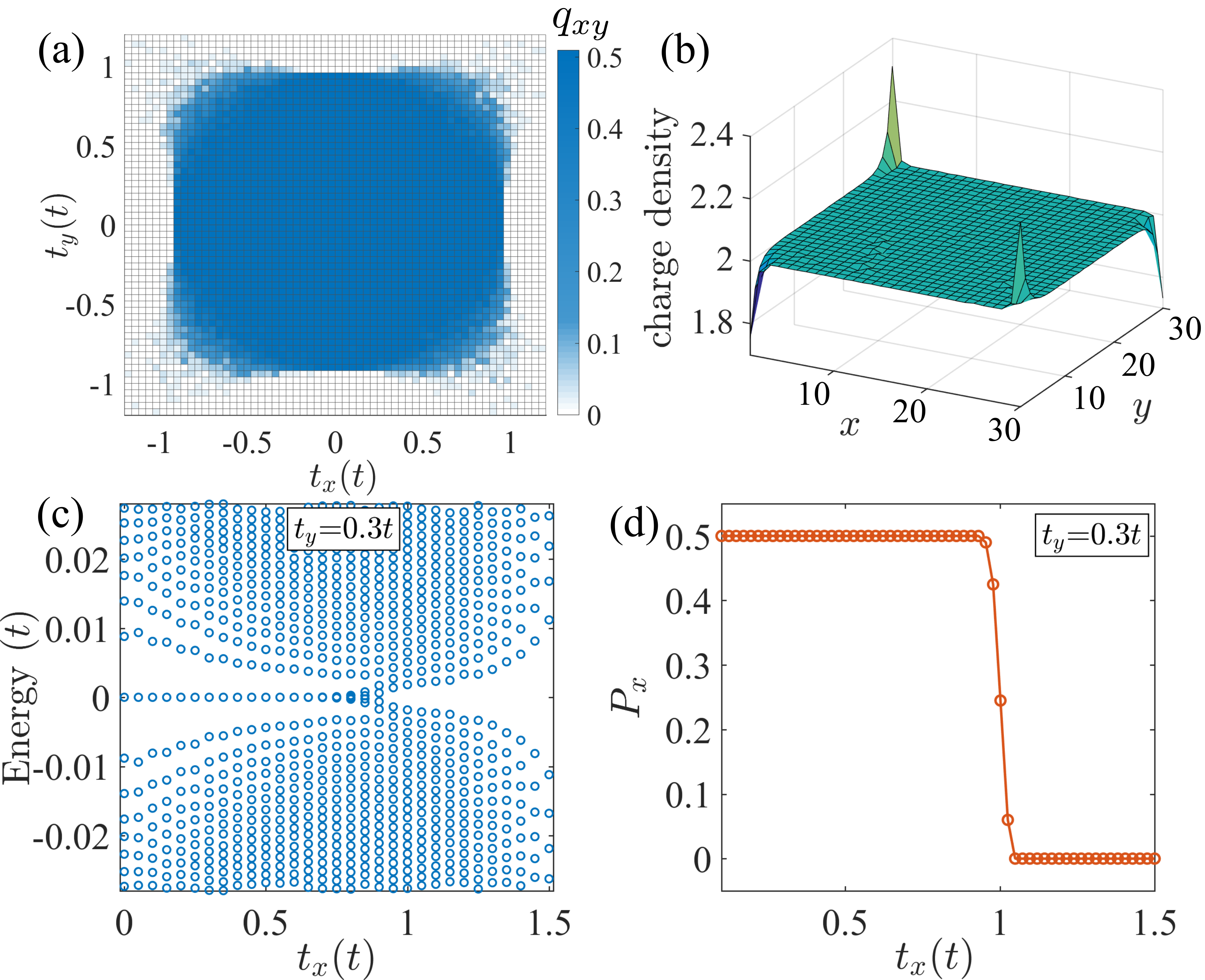}

\caption{(a) Phase diagram of $q_{xy}$ against $t_{x}$ and $t_{y}$. The
dimension of the system is $L=30$ with 30 random-flux
configurations. (b) Electron charge density in the
extrinsic HOTI phase at half-filling.  (c) Disorder-averaged energy spectrum as a function of $t_{x}$ for $t_{y}=0.3t$ under open boundary conditions. (d) Disorder-averaged edge polarization $P_{x}$ as a function of $t_{x}$ for $t_{y}=0.3t$. $U=2\pi$ for all plots.
\label{fig:HOTI}}
\end{figure}

\textit{Extrinsic HOTI induced by random flux.}---Now, we show that in the parameter regime $|t_x|<t$ and  $|t_y|<t$, the  band insulator induced by random flux can be an extrinsic HOTI ~\cite{Geier18prb, Trifunovic19prx}. For concreteness, we consider $U=2\pi$. In this case, the system is an insulator with a finite energy gap, unless $|t_{x}|=|t_{y}|= t$, \textit{c.f.} Fig.\ \ref{fig:lattice}(d)~\cite{Note4}. Note that the disorder-averaged flux on each plaquette is zero. The defined electric quadrupole moment $q_{xy}$ can provide a topological index to characterize the extrinsic HOTI~\cite{Benalcazar17Science,BBH17prb,Kang19prb,Wheeler19prb,Roy19prr}. In the phase diagram shown in Fig.~\ref{fig:HOTI}(a), which is similar to that of BBH model, we observe a
nontrivial region (blue) with a half quantized $q_{xy}=1/2$. In the outer
region (white), the system is a trivial insulator with $q_{xy}=0$. This implies that the random-flux driven higher-order topological phases can be continuously connected to that of the BBH model.
The quantization of $q_{xy}$ is protected by chiral symmetry~\cite{LiCA20prl}. Accordingly, a nontrivial $q_{xy}$ indicates the emergence of zero-energy modes at the corners of the system. This is confirmed numerically in Figs.~\ref{fig:lattice}(c)
and \ref{fig:HOTI}(c) where four zero-energy modes clearly emerge
in the nontrivial phase whereas they disappear in the trivial phase.
Furthermore, we calculate the local charge density at half-filling
[Fig.~\ref{fig:HOTI}(b)]. Summing the charge density over
each quadrant including a single corner, we find that the total charge
takes fractional values $\pm1/2$ as long as $L$ is large enough. These fractional corner charges provide another hallmark of the HOTI.

For a fixed strong $U$, the system transits between
an extrinsic HOTI and a trivial insulator by changing $t_{x}$ or $t_{y}$. Due to its extrinsic nature, the topological phase transitions take place at the boundaries instead of the bulk of the system. To elucidate this phase transition,
we calculate the effective Hamiltonian $H_{\mathrm{edge}}$ for edges
in the presence of random flux via a recursive Green's function
method~\cite{PengY17prb,Note5}. We see the edge spectrum close and reopen around phase boundary. Alternatively, the transition can also be shown from the edge polarization of $H_{\mathrm{edge}}$~\cite{BBH17prb,Resta98prl}. For illustration, we consider the edge along $x$-direction and present the disorder-averaged
polarization $P_{x}$ as a function of $t_{x}$ in Fig.\ \ref{fig:HOTI}(d).
Near $t_{x}=t$, $P_{x}$ changes suddenly from $1/2$ to 0, indicating
a topological phase transition. The results for edges along $y$-direction
can be obtained similarly. We note that the system is nontrivial
only if both edge Hamiltonians along $x$- and $y$-directions are
nontrivial.

\textit{Discussion and conclusions.}---
Note that the metal-band insulator transition driven by random flux is found to also occur in the topologically trivial regime~\cite{Li2021SM}, which indicates its generality. Clearly, the random flux with zero mean is different from the case with a uniform
flux, where the Hofstadter butterfly emerges~\cite{Otaki19prb,zuo21arxiv,Hofstadter76prb}. In the limit of $t_x=t_y=t$, our system reduces to the conventional random-flux model.
In this limit, we recover the well established result that the bulk states at the band center stay delocalized and no metal-insulator transition occurs~\cite{Li2021SM}.  We emphasize that the random-flux driven metal-band insulator transition is distinctively different from related work in interacting systems~\cite{AnJ01prb,Foster08prb} where the competition between (random) flux and electron-electron interaction is responsible for an interaction driven phase transition.

The 2D SSH model can be realized in different platforms such as metamaterials~\cite{Serra-Garcia18nature, Ni19nm, XieBY19prl,QiY20prl,ChenXD19prl}, microwave and electric circuits~\cite{Peterson18nature,Imhof18np,Roy21prr,ZhangW21prl}. In particular, the manipulation of effective magnetic fluxes has become experimentally accessible in sonic crystals and circuit simulators~\cite{Linzk21arxiv,LiS21arxiv}. Therefore, these materials may provide us with promising platforms to test our predictions by taking advantage of their high controllability.

In conclusion, based on the 2D SSH model we have revealed the first example of a metal-band insulator transition that is solely driven by random flux.
We have analyzed this metal-band insulator transition by level statistics
and finite-size scaling theory, and found the critical exponent as
$\nu=2.48\pm0.08$. It is shown that the emergent insulator
can be an extrinsic HOTI by presenting its phase diagram and characteristic boundary
signatures. We have further proposed an effective band structure picture
to understand the metal-band insulator transition driven by random flux.

\begin{acknowledgements}This work was supported by the DFG (SPP1666,
SFB1170 ``ToCoTronics'', and SFB1143), the W\"urzburg-Dresden Cluster of Excellence
ct.qmat, EXC2147, Project-id 390858490, and the Elitenetzwerk Bayern
Graduate School on ``Topological Insulators''. C.A. L. thanks Bo Fu and Jian Li for helpful discussions. \end{acknowledgements}

%\bibliographystyle{apsrev4-1}
%\bibliographystyle{apsrev4-1-etal-title}

% \bibliography{Refsdata}

%merlin.mbs apsrev4-1.bst 2010-07-25 4.21a (PWD, AO, DPC) hacked
%Control: key (0)
%Control: author (72) initials jnrlst
%Control: editor formatted (1) identically to author
%Control: production of article title (-1) disabled
%Control: page (0) single
%Control: year (1) truncated
%Control: production of eprint (0) enabled
\begin{thebibliography}{0}%
\makeatletter
\providecommand \@ifxundefined [1]{%
 \@ifx{#1\undefined}
}%
\providecommand \@ifnum [1]{%
 \ifnum #1\expandafter \@firstoftwo
 \else \expandafter \@secondoftwo
 \fi
}%
\providecommand \@ifx [1]{%
 \ifx #1\expandafter \@firstoftwo
 \else \expandafter \@secondoftwo
 \fi
}%
\providecommand \natexlab [1]{#1}%
\providecommand \enquote  [1]{``#1''}%
\providecommand \bibnamefont  [1]{#1}%
\providecommand \bibfnamefont [1]{#1}%
\providecommand \citenamefont [1]{#1}%
\providecommand \href@noop [0]{\@secondoftwo}%
\providecommand \href [0]{\begingroup \@sanitize@url \@href}%
\providecommand \@href[1]{\@@startlink{#1}\@@href}%
\providecommand \@@href[1]{\endgroup#1\@@endlink}%
\providecommand \@sanitize@url [0]{\catcode `\\12\catcode `\$12\catcode
  `\&12\catcode `\#12\catcode `\^12\catcode `\_12\catcode `\%12\relax}%
\providecommand \@@startlink[1]{}%
\providecommand \@@endlink[0]{}%
\providecommand \url  [0]{\begingroup\@sanitize@url \@url }%
\providecommand \@url [1]{\endgroup\@href {#1}{\urlprefix }}%
\providecommand \urlprefix  [0]{URL }%
\providecommand \Eprint [0]{\href }%
\providecommand \doibase [0]{http://dx.doi.org/}%
\providecommand \selectlanguage [0]{\@gobble}%
\providecommand \bibinfo  [0]{\@secondoftwo}%
\providecommand \bibfield  [0]{\@secondoftwo}%
\providecommand \translation [1]{[#1]}%
\providecommand \BibitemOpen [0]{}%
\providecommand \bibitemStop [0]{}%
\providecommand \bibitemNoStop [0]{.\EOS\space}%
\providecommand \EOS [0]{\spacefactor3000\relax}%
\providecommand \BibitemShut  [1]{\csname bibitem#1\endcsname}%
\let\auto@bib@innerbib\@empty
%</preamble>
\end{thebibliography}%


\begin{thebibliography}{89}%
\makeatletter
\providecommand \@ifxundefined [1]{%
 \@ifx{#1\undefined}
}%
\providecommand \@ifnum [1]{%
 \ifnum #1\expandafter \@firstoftwo
 \else \expandafter \@secondoftwo
 \fi
}%
\providecommand \@ifx [1]{%
 \ifx #1\expandafter \@firstoftwo
 \else \expandafter \@secondoftwo
 \fi
}%
\providecommand \natexlab [1]{#1}%
\providecommand \enquote  [1]{``#1''}%
\providecommand \bibnamefont  [1]{#1}%
\providecommand \bibfnamefont [1]{#1}%
\providecommand \citenamefont [1]{#1}%
\providecommand \href@noop [0]{\@secondoftwo}%
\providecommand \href [0]{\begingroup \@sanitize@url \@href}%
\providecommand \@href[1]{\@@startlink{#1}\@@href}%
\providecommand \@@href[1]{\endgroup#1\@@endlink}%
\providecommand \@sanitize@url [0]{\catcode `\\12\catcode `\$12\catcode
  `\&12\catcode `\#12\catcode `\^12\catcode `\_12\catcode `\%12\relax}%
\providecommand \@@startlink[1]{}%
\providecommand \@@endlink[0]{}%
\providecommand \url  [0]{\begingroup\@sanitize@url \@url }%
\providecommand \@url [1]{\endgroup\@href {#1}{\urlprefix }}%
\providecommand \urlprefix  [0]{URL }%
\providecommand \Eprint [0]{\href }%
\providecommand \doibase [0]{http://dx.doi.org/}%
\providecommand \selectlanguage [0]{\@gobble}%
\providecommand \bibinfo  [0]{\@secondoftwo}%
\providecommand \bibfield  [0]{\@secondoftwo}%
\providecommand \translation [1]{[#1]}%
\providecommand \BibitemOpen [0]{}%
\providecommand \bibitemStop [0]{}%
\providecommand \bibitemNoStop [0]{.\EOS\space}%
\providecommand \EOS [0]{\spacefactor3000\relax}%
\providecommand \BibitemShut  [1]{\csname bibitem#1\endcsname}%
\let\auto@bib@innerbib\@empty
%</preamble>
\bibitem [{\citenamefont {Anderson}(1958)}]{Anderson58pr}%
  \BibitemOpen
  \bibfield  {author} {\bibinfo {author} {\bibfnamefont {P.~W.}\ \bibnamefont
  {Anderson}},\ }\bibfield  {title} {\enquote {\bibinfo {title} {Absence of
  diffusion in certain random lattices}}, }\href {\doibase
  10.1103/PhysRev.109.1492} {\bibfield  {journal} {\bibinfo  {journal} {Phys.
  Rev.}\ }\textbf {\bibinfo {volume} {109}},\ \bibinfo {pages} {1492} (\bibinfo
  {year} {1958})}\BibitemShut {NoStop}%
\bibitem [{\citenamefont {Evers}\ and\ \citenamefont
  {Mirlin}(2008)}]{Evers08rmp}%
  \BibitemOpen
  \bibfield  {author} {\bibinfo {author} {\bibfnamefont {F.}~\bibnamefont
  {Evers}}\ and\ \bibinfo {author} {\bibfnamefont {A.~D.}\ \bibnamefont
  {Mirlin}},\ }\bibfield  {title} {\enquote {\bibinfo {title} {Anderson
  transitions}}, }\href {\doibase 10.1103/RevModPhys.80.1355} {\bibfield
  {journal} {\bibinfo  {journal} {Rev. Mod. Phys.}\ }\textbf {\bibinfo {volume}
  {80}},\ \bibinfo {pages} {1355} (\bibinfo {year} {2008})}\BibitemShut
  {NoStop}%
\bibitem [{\citenamefont {Sanchez-Palencia}\ and\ \citenamefont
  {Lewenstein}(2010)}]{Sanchez-Palencia10NP}%
  \BibitemOpen
  \bibfield  {author} {\bibinfo {author} {\bibfnamefont {L.}~\bibnamefont
  {Sanchez-Palencia}}\ and\ \bibinfo {author} {\bibfnamefont {M.}~\bibnamefont
  {Lewenstein}},\ }\bibfield  {title} {\enquote {\bibinfo {title} {Disordered
  quantum gases under control}}, }\href {\doibase 10.1038/nphys1507} {\bibfield
   {journal} {\bibinfo  {journal} {Nature Physics}\ }\textbf {\bibinfo {volume}
  {6}},\ \bibinfo {pages} {87} (\bibinfo {year} {2010})}\BibitemShut {NoStop}%
\bibitem [{\citenamefont {Schwartz}\ \emph {et~al.}(2007)\citenamefont
  {Schwartz}, \citenamefont {Bartal}, \citenamefont {Fishman},\ and\
  \citenamefont {Segev}}]{Schwartz07nature}%
  \BibitemOpen
  \bibfield  {author} {\bibinfo {author} {\bibfnamefont {T.}~\bibnamefont
  {Schwartz}}, \bibinfo {author} {\bibfnamefont {G.}~\bibnamefont {Bartal}},
  \bibinfo {author} {\bibfnamefont {S.}~\bibnamefont {Fishman}}, \ and\
  \bibinfo {author} {\bibfnamefont {M.}~\bibnamefont {Segev}},\ }\bibfield
  {title} {\enquote {\bibinfo {title} {Transport and Anderson localization in
  disordered two-dimensional photonic lattices}}, }\href {\doibase
  10.1038/nature05623} {\bibfield  {journal} {\bibinfo  {journal} {Nature}\
  }\textbf {\bibinfo {volume} {446}},\ \bibinfo {pages} {52} (\bibinfo {year}
  {2007})}\BibitemShut {NoStop}%
\bibitem [{\citenamefont {Chab\'e}\ \emph {et~al.}(2008)\citenamefont
  {Chab\'e}, \citenamefont {Lemari\'e}, \citenamefont {Gr\'emaud},
  \citenamefont {Delande}, \citenamefont {Szriftgiser},\ and\ \citenamefont
  {Garreau}}]{Chabe08prl}%
  \BibitemOpen
  \bibfield  {author} {\bibinfo {author} {\bibfnamefont {J.}~\bibnamefont
  {Chab\'e}}, \bibinfo {author} {\bibfnamefont {G.}~\bibnamefont {Lemari\'e}},
  \bibinfo {author} {\bibfnamefont {B.}~\bibnamefont {Gr\'emaud}}, \bibinfo
  {author} {\bibfnamefont {D.}~\bibnamefont {Delande}}, \bibinfo {author}
  {\bibfnamefont {P.}~\bibnamefont {Szriftgiser}}, \ and\ \bibinfo {author}
  {\bibfnamefont {J.~C.}\ \bibnamefont {Garreau}},\ }\bibfield  {title}
  {\enquote {\bibinfo {title} {Experimental observation of the Anderson
  metal-insulator transition with atomic matter waves}}, }\href {\doibase
  10.1103/PhysRevLett.101.255702} {\bibfield  {journal} {\bibinfo  {journal}
  {Phys. Rev. Lett.}\ }\textbf {\bibinfo {volume} {101}},\ \bibinfo {pages}
  {255702} (\bibinfo {year} {2008})}\BibitemShut {NoStop}%
\bibitem [{\citenamefont {Li}\ \emph {et~al.}(2009)\citenamefont {Li},
  \citenamefont {Chu}, \citenamefont {Jain},\ and\ \citenamefont
  {Shen}}]{LiJian09prl}%
  \BibitemOpen
  \bibfield  {author} {\bibinfo {author} {\bibfnamefont {J.}~\bibnamefont
  {Li}}, \bibinfo {author} {\bibfnamefont {R.-L.}\ \bibnamefont {Chu}},
  \bibinfo {author} {\bibfnamefont {J.~K.}\ \bibnamefont {Jain}}, \ and\
  \bibinfo {author} {\bibfnamefont {S.-Q.}\ \bibnamefont {Shen}},\ }\bibfield
  {title} {\enquote {\bibinfo {title} {Topological Anderson insulator}}, }\href
  {\doibase 10.1103/PhysRevLett.102.136806} {\bibfield  {journal} {\bibinfo
  {journal} {Phys. Rev. Lett.}\ }\textbf {\bibinfo {volume} {102}},\ \bibinfo
  {pages} {136806} (\bibinfo {year} {2009})}\BibitemShut {NoStop}%
\bibitem [{\citenamefont {Groth}\ \emph {et~al.}(2009)\citenamefont {Groth},
  \citenamefont {Wimmer}, \citenamefont {Akhmerov}, \citenamefont
  {Tworzyd\l{}o},\ and\ \citenamefont {Beenakker}}]{Groth09prl}%
  \BibitemOpen
  \bibfield  {author} {\bibinfo {author} {\bibfnamefont {C.~W.}\ \bibnamefont
  {Groth}}, \bibinfo {author} {\bibfnamefont {M.}~\bibnamefont {Wimmer}},
  \bibinfo {author} {\bibfnamefont {A.~R.}\ \bibnamefont {Akhmerov}}, \bibinfo
  {author} {\bibfnamefont {J.}~\bibnamefont {Tworzyd\l{}o}}, \ and\ \bibinfo
  {author} {\bibfnamefont {C.~W.~J.}\ \bibnamefont {Beenakker}},\ }\bibfield
  {title} {\enquote {\bibinfo {title} {Theory of the topological Anderson
  insulator}}, }\href {\doibase 10.1103/PhysRevLett.103.196805} {\bibfield
  {journal} {\bibinfo  {journal} {Phys. Rev. Lett.}\ }\textbf {\bibinfo
  {volume} {103}},\ \bibinfo {pages} {196805} (\bibinfo {year}
  {2009})}\BibitemShut {NoStop}%
\bibitem [{\citenamefont {Jiang}\ \emph {et~al.}(2009)\citenamefont {Jiang},
  \citenamefont {Wang}, \citenamefont {Sun},\ and\ \citenamefont
  {Xie}}]{JiangH09prb}%
  \BibitemOpen
  \bibfield  {author} {\bibinfo {author} {\bibfnamefont {H.}~\bibnamefont
  {Jiang}}, \bibinfo {author} {\bibfnamefont {L.}~\bibnamefont {Wang}},
  \bibinfo {author} {\bibfnamefont {Q.-F.}\ \bibnamefont {Sun}}, \ and\
  \bibinfo {author} {\bibfnamefont {X.~C.}\ \bibnamefont {Xie}},\ }\bibfield
  {title} {\enquote {\bibinfo {title} {Numerical study of the topological
 Anderson insulator in HgTe/CdTe quantum wells}}, }\href {\doibase
  10.1103/PhysRevB.80.165316} {\bibfield  {journal} {\bibinfo  {journal} {Phys.
  Rev. B}\ }\textbf {\bibinfo {volume} {80}},\ \bibinfo {pages} {165316}
  (\bibinfo {year} {2009})}\BibitemShut {NoStop}%
\bibitem [{\citenamefont {Guo}\ \emph {et~al.}(2010)\citenamefont {Guo},
  \citenamefont {Rosenberg}, \citenamefont {Refael},\ and\ \citenamefont
  {Franz}}]{GuoHM10prl}%
  \BibitemOpen
  \bibfield  {author} {\bibinfo {author} {\bibfnamefont {H.-M.}\ \bibnamefont
  {Guo}}, \bibinfo {author} {\bibfnamefont {G.}~\bibnamefont {Rosenberg}},
  \bibinfo {author} {\bibfnamefont {G.}~\bibnamefont {Refael}}, \ and\ \bibinfo
  {author} {\bibfnamefont {M.}~\bibnamefont {Franz}},\ }\bibfield  {title}
  {\enquote {\bibinfo {title} {Topological Anderson insulator in three
  dimensions}}, }\href {\doibase 10.1103/PhysRevLett.105.216601} {\bibfield
  {journal} {\bibinfo  {journal} {Phys. Rev. Lett.}\ }\textbf {\bibinfo
  {volume} {105}},\ \bibinfo {pages} {216601} (\bibinfo {year}
  {2010})}\BibitemShut {NoStop}%
\bibitem [{\citenamefont {Meier}\ \emph {et~al.}(2018)\citenamefont {Meier},
  \citenamefont {An}, \citenamefont {Dauphin}, \citenamefont {Maffei},
  \citenamefont {Massignan}, \citenamefont {Hughes},\ and\ \citenamefont
  {Gadway}}]{Meier19Science}%
  \BibitemOpen
  \bibfield  {author} {\bibinfo {author} {\bibfnamefont {E.~J.}\ \bibnamefont
  {Meier}}, \bibinfo {author} {\bibfnamefont {F.~A.}\ \bibnamefont {An}},
  \bibinfo {author} {\bibfnamefont {A.}~\bibnamefont {Dauphin}}, \bibinfo
  {author} {\bibfnamefont {M.}~\bibnamefont {Maffei}}, \bibinfo {author}
  {\bibfnamefont {P.}~\bibnamefont {Massignan}}, \bibinfo {author}
  {\bibfnamefont {T.~L.}\ \bibnamefont {Hughes}}, \ and\ \bibinfo {author}
  {\bibfnamefont {B.}~\bibnamefont {Gadway}},\ }\bibfield  {title} {\enquote
  {\bibinfo {title} {Observation of the topological Anderson insulator in
  disordered atomic wires}}, }\href {\doibase 10.1126/science.aat3406}
  {\bibfield  {journal} {\bibinfo  {journal} {Science}\ }\textbf {\bibinfo
  {volume} {362}},\ \bibinfo {pages} {929} (\bibinfo {year}
  {2018})}\BibitemShut {NoStop}%
\bibitem [{\citenamefont {St{\"u}tzer}\ \emph {et~al.}(2018)\citenamefont
  {St{\"u}tzer}, \citenamefont {Plotnik}, \citenamefont {Lumer}, \citenamefont
  {Titum}, \citenamefont {Lindner}, \citenamefont {Segev}, \citenamefont
  {Rechtsman},\ and\ \citenamefont {Szameit}}]{Stutzer18nature}%
  \BibitemOpen
  \bibfield  {author} {\bibinfo {author} {\bibfnamefont {S.}~\bibnamefont
  {St{\"u}tzer}}, \bibinfo {author} {\bibfnamefont {Y.}~\bibnamefont
  {Plotnik}}, \bibinfo {author} {\bibfnamefont {Y.}~\bibnamefont {Lumer}},
  \bibinfo {author} {\bibfnamefont {P.}~\bibnamefont {Titum}}, \bibinfo
  {author} {\bibfnamefont {N.~H.}\ \bibnamefont {Lindner}}, \bibinfo {author}
  {\bibfnamefont {M.}~\bibnamefont {Segev}}, \bibinfo {author} {\bibfnamefont
  {M.~C.}\ \bibnamefont {Rechtsman}}, \ and\ \bibinfo {author} {\bibfnamefont
  {A.}~\bibnamefont {Szameit}},\ }\bibfield  {title} {\enquote {\bibinfo
  {title} {Photonic topological Anderson insulators}}, }\href {\doibase
  10.1038/s41586-018-0418-2} {\bibfield  {journal} {\bibinfo  {journal}
  {Nature}\ }\textbf {\bibinfo {volume} {560}},\ \bibinfo {pages} {461}
  (\bibinfo {year} {2018})}\BibitemShut {NoStop}%
\bibitem [{\citenamefont {Cerovski}(2001)}]{Cerovski01prb}%
  \BibitemOpen
  \bibfield  {author} {\bibinfo {author} {\bibfnamefont {V.~Z.}\ \bibnamefont
  {Cerovski}},\ }\bibfield  {title} {\enquote {\bibinfo {title} {Critical
  exponent of the random flux model on an infinite two-dimensional square
  lattice and anomalous critical states}}, }\href {\doibase
  10.1103/PhysRevB.64.161101} {\bibfield  {journal} {\bibinfo  {journal} {Phys.
  Rev. B}\ }\textbf {\bibinfo {volume} {64}},\ \bibinfo {pages} {161101(R)}
  (\bibinfo {year} {2001})}\BibitemShut {NoStop}%
\bibitem [{\citenamefont {Furusaki}(1999)}]{Furusaki99prl}%
  \BibitemOpen
  \bibfield  {author} {\bibinfo {author} {\bibfnamefont {A.}~\bibnamefont
  {Furusaki}},\ }\bibfield  {title} {\enquote {\bibinfo {title} {Anderson
  localization due to a random magnetic field in two dimensions}}, }\href
  {\doibase 10.1103/PhysRevLett.82.604} {\bibfield  {journal} {\bibinfo
  {journal} {Phys. Rev. Lett.}\ }\textbf {\bibinfo {volume} {82}},\ \bibinfo
  {pages} {604} (\bibinfo {year} {1999})}\BibitemShut {NoStop}%
\bibitem [{\citenamefont {Aronov}\ \emph {et~al.}(1994)\citenamefont {Aronov},
  \citenamefont {Mirlin},\ and\ \citenamefont {W\"olfle}}]{Aronov94prb}%
  \BibitemOpen
  \bibfield  {author} {\bibinfo {author} {\bibfnamefont {A.~G.}\ \bibnamefont
  {Aronov}}, \bibinfo {author} {\bibfnamefont {A.~D.}\ \bibnamefont {Mirlin}},
  \ and\ \bibinfo {author} {\bibfnamefont {P.}~\bibnamefont {W\"olfle}},\
  }\bibfield  {title} {\enquote {\bibinfo {title} {Localization of charged
  quantum particles in a static random magnetic field}}, }\href {\doibase
  10.1103/PhysRevB.49.16609} {\bibfield  {journal} {\bibinfo  {journal} {Phys.
  Rev. B}\ }\textbf {\bibinfo {volume} {49}},\ \bibinfo {pages} {16609}
  (\bibinfo {year} {1994})}\BibitemShut {NoStop}%
\bibitem [{\citenamefont {Taras-Semchuk}\ and\ \citenamefont
  {Efetov}(2000)}]{Taras00prl}%
  \BibitemOpen
  \bibfield  {author} {\bibinfo {author} {\bibfnamefont {D.}~\bibnamefont
  {Taras-Semchuk}}\ and\ \bibinfo {author} {\bibfnamefont {K.~B.}\ \bibnamefont
  {Efetov}},\ }\bibfield  {title} {\enquote {\bibinfo {title} {Antilocalization
  in a 2d electron gas in a random magnetic field}}, }\href {\doibase
  10.1103/PhysRevLett.85.1060} {\bibfield  {journal} {\bibinfo  {journal}
  {Phys. Rev. Lett.}\ }\textbf {\bibinfo {volume} {85}},\ \bibinfo {pages}
  {1060} (\bibinfo {year} {2000})}\BibitemShut {NoStop}%
\bibitem [{\citenamefont {Sheng}\ and\ \citenamefont
  {Weng}(1995)}]{Sheng95prl}%
  \BibitemOpen
  \bibfield  {author} {\bibinfo {author} {\bibfnamefont {D.~N.}\ \bibnamefont
  {Sheng}}\ and\ \bibinfo {author} {\bibfnamefont {Z.~Y.}\ \bibnamefont
  {Weng}},\ }\bibfield  {title} {\enquote {\bibinfo {title} {Delocalization of
  electrons in a random magnetic field}}, }\href {\doibase
  10.1103/PhysRevLett.75.2388} {\bibfield  {journal} {\bibinfo  {journal}
  {Phys. Rev. Lett.}\ }\textbf {\bibinfo {volume} {75}},\ \bibinfo {pages}
  {2388} (\bibinfo {year} {1995})}\BibitemShut {NoStop}%
\bibitem [{\citenamefont {Liu}\ \emph {et~al.}(1995)\citenamefont {Liu},
  \citenamefont {Xie}, \citenamefont {Das~Sarma},\ and\ \citenamefont
  {Zhang}}]{LiuDZ95prb}%
  \BibitemOpen
  \bibfield  {author} {\bibinfo {author} {\bibfnamefont {D.~Z.}\ \bibnamefont
  {Liu}}, \bibinfo {author} {\bibfnamefont {X.~C.}\ \bibnamefont {Xie}},
  \bibinfo {author} {\bibfnamefont {S.}~\bibnamefont {Das~Sarma}}, \ and\
  \bibinfo {author} {\bibfnamefont {S.~C.}\ \bibnamefont {Zhang}},\ }\bibfield
  {title} {\enquote {\bibinfo {title} {Electron localization in a
  two-dimensional system with random magnetic flux}}, }\href {\doibase
  10.1103/PhysRevB.52.5858} {\bibfield  {journal} {\bibinfo  {journal} {Phys.
  Rev. B}\ }\textbf {\bibinfo {volume} {52}},\ \bibinfo {pages} {5858}
  (\bibinfo {year} {1995})}\BibitemShut {NoStop}%
\bibitem [{\citenamefont {Zhang}\ and\ \citenamefont
  {Arovas}(1994)}]{ZhangSC94prl}%
  \BibitemOpen
  \bibfield  {author} {\bibinfo {author} {\bibfnamefont {S.-C.}\ \bibnamefont
  {Zhang}}\ and\ \bibinfo {author} {\bibfnamefont {D.~P.}\ \bibnamefont
  {Arovas}},\ }\bibfield  {title} {\enquote {\bibinfo {title} {Effective field
  theory of electron motion in the presence of random magnetic flux}}, }\href
  {\doibase 10.1103/PhysRevLett.72.1886} {\bibfield  {journal} {\bibinfo
  {journal} {Phys. Rev. Lett.}\ }\textbf {\bibinfo {volume} {72}},\ \bibinfo
  {pages} {1886} (\bibinfo {year} {1994})}\BibitemShut {NoStop}%
\bibitem [{\citenamefont {Gade}(1993)}]{Gade93npb}%
  \BibitemOpen
  \bibfield  {author} {\bibinfo {author} {\bibfnamefont {R.}~\bibnamefont
  {Gade}},\ }\bibfield  {title} {\enquote {\bibinfo {title} {Anderson
  localization for sublattice models}}, }\href {\doibase
  https://doi.org/10.1016/0550-3213(93)90601-K} {\bibfield  {journal} {\bibinfo
   {journal} {Nuclear Physics B}\ }\textbf {\bibinfo {volume} {398}},\ \bibinfo
  {pages} {499} (\bibinfo {year} {1993})}\BibitemShut {NoStop}%
\bibitem [{\citenamefont {Sugiyama}\ and\ \citenamefont
  {Nagaosa}(1993)}]{Sugiyama93prl}%
  \BibitemOpen
  \bibfield  {author} {\bibinfo {author} {\bibfnamefont {T.}~\bibnamefont
  {Sugiyama}}\ and\ \bibinfo {author} {\bibfnamefont {N.}~\bibnamefont
  {Nagaosa}},\ }\bibfield  {title} {\enquote {\bibinfo {title} {Localization in
  a random magnetic field in 2d}}, }\href {\doibase
  10.1103/PhysRevLett.70.1980} {\bibfield  {journal} {\bibinfo  {journal}
  {Phys. Rev. Lett.}\ }\textbf {\bibinfo {volume} {70}},\ \bibinfo {pages}
  {1980} (\bibinfo {year} {1993})}\BibitemShut {NoStop}%
\bibitem [{\citenamefont {Avishai}\ \emph {et~al.}(1993)\citenamefont
  {Avishai}, \citenamefont {Hatsugai},\ and\ \citenamefont
  {Kohmoto}}]{Avishai93prb}%
  \BibitemOpen
  \bibfield  {author} {\bibinfo {author} {\bibfnamefont {Y.}~\bibnamefont
  {Avishai}}, \bibinfo {author} {\bibfnamefont {Y.}~\bibnamefont {Hatsugai}}, \
  and\ \bibinfo {author} {\bibfnamefont {M.}~\bibnamefont {Kohmoto}},\
  }\bibfield  {title} {\enquote {\bibinfo {title} {Localization problem of a
  two-dimensional lattice in a random magnetic field}}, }\href {\doibase
  10.1103/PhysRevB.47.9561} {\bibfield  {journal} {\bibinfo  {journal} {Phys.
  Rev. B}\ }\textbf {\bibinfo {volume} {47}},\ \bibinfo {pages} {9561}
  (\bibinfo {year} {1993})}\BibitemShut {NoStop}%
\bibitem [{\citenamefont {Lee}\ and\ \citenamefont
  {Fisher}(1981)}]{PALee81prl}%
  \BibitemOpen
  \bibfield  {author} {\bibinfo {author} {\bibfnamefont {P.~A.}\ \bibnamefont
  {Lee}}\ and\ \bibinfo {author} {\bibfnamefont {D.~S.}\ \bibnamefont
  {Fisher}},\ }\bibfield  {title} {\enquote {\bibinfo {title} {Anderson
  localization in two dimensions}}, }\href {\doibase
  10.1103/PhysRevLett.47.882} {\bibfield  {journal} {\bibinfo  {journal} {Phys.
  Rev. Lett.}\ }\textbf {\bibinfo {volume} {47}},\ \bibinfo {pages} {882}
  (\bibinfo {year} {1981})}\BibitemShut {NoStop}%
\bibitem [{\citenamefont {AnJ}\ and\ \citenamefont
  {Lin}(2001)}]{AnJ01prb}%
  \BibitemOpen
  \bibfield  {author} {\bibinfo {author} {\bibfnamefont {J.}\ \bibnamefont
  {An}},\ \bibinfo {author} {\bibfnamefont {C. D.}\ \bibnamefont
  {Gong}}\ and\ \bibinfo {author} {\bibfnamefont {H. Q.}\ \bibnamefont
  {Lin}},\ }\bibfield  {title} {\enquote {\bibinfo {title} {Theory of the magnetic-field-induced metal-insulator transition}}, }\href {\doibase
  10.1103/PhysRevB.63.174434} {\bibfield  {journal} {\bibinfo  {journal} {Phys.
  Rev. B}\ }\textbf {\bibinfo {volume} {63}},\ \bibinfo {pages} {174434}
  (\bibinfo {year} {2001})}\BibitemShut {NoStop}%
 \bibitem [{\citenamefont {Foster}\ and\ \citenamefont
  {Ludwig}(2008)}]{Foster08prb}%
  \BibitemOpen
  \bibfield  {author} {\bibinfo {author} {\bibfnamefont {M. S.}\ \bibnamefont
  {Foster}}\ and\ \bibinfo {author} {\bibfnamefont {A. W. W.}\ \bibnamefont
  {Ludwig}},\ }\bibfield  {title} {\enquote {\bibinfo {title} {Metal-insulator transition from combined disorder and interaction effects in Hubbard-like electronic lattice models with random hopping}}, }\href {\doibase
 10.1103/PhysRevB.77.165108} {\bibfield  {journal} {\bibinfo  {journal} {Phys.
  Rev. B}\ }\textbf {\bibinfo {volume} {77}},\ \bibinfo {pages} {165108}
  (\bibinfo {year} {2008})}\BibitemShut {NoStop}%
\bibitem [{\citenamefont {Liu}\ and\ \citenamefont
  {Wakabayashi}(2017)}]{LiuF17prl}%
  \BibitemOpen
  \bibfield  {author} {\bibinfo {author} {\bibfnamefont {F.}~\bibnamefont
  {Liu}}\ and\ \bibinfo {author} {\bibfnamefont {K.}~\bibnamefont
  {Wakabayashi}},\ }\bibfield  {title} {\enquote {\bibinfo {title} {Novel
  topological phase with a zero Berry curvature}}, }\href {\doibase
  10.1103/PhysRevLett.118.076803} {\bibfield  {journal} {\bibinfo  {journal}
  {Phys. Rev. Lett.}\ }\textbf {\bibinfo {volume} {118}},\ \bibinfo {pages}
  {076803} (\bibinfo {year} {2017})}\BibitemShut {NoStop}%
\bibitem [{\citenamefont {Benalcazar}\ \emph
  {et~al.}(2017{\natexlab{a}})\citenamefont {Benalcazar}, \citenamefont
  {Bernevig},\ and\ \citenamefont {Hughes}}]{Benalcazar17Science}%
  \BibitemOpen
  \bibfield  {author} {\bibinfo {author} {\bibfnamefont {W.~A.}\ \bibnamefont
  {Benalcazar}}, \bibinfo {author} {\bibfnamefont {B.~A.}\ \bibnamefont
  {Bernevig}}, \ and\ \bibinfo {author} {\bibfnamefont {T.~L.}\ \bibnamefont
  {Hughes}},\ }\bibfield  {title} {\enquote {\bibinfo {title} {Quantized
  electric multipole insulators}}, }\href {\doibase 10.1126/science.aah6442}
  {\bibfield  {journal} {\bibinfo  {journal} {Science}\ }\textbf {\bibinfo
  {volume} {357}},\ \bibinfo {pages} {61} (\bibinfo {year}
  {2017}{\natexlab{a}})}\BibitemShut {NoStop}%
\bibitem [{\citenamefont {Benalcazar}\ \emph
  {et~al.}(2017{\natexlab{b}})\citenamefont {Benalcazar}, \citenamefont
  {Bernevig},\ and\ \citenamefont {Hughes}}]{BBH17prb}%
  \BibitemOpen
  \bibfield  {author} {\bibinfo {author} {\bibfnamefont {W.~A.}\ \bibnamefont
  {Benalcazar}}, \bibinfo {author} {\bibfnamefont {B.~A.}\ \bibnamefont
  {Bernevig}}, \ and\ \bibinfo {author} {\bibfnamefont {T.~L.}\ \bibnamefont
  {Hughes}},\ }\bibfield  {title} {\enquote {\bibinfo {title} {Electric
  multipole moments, topological multipole moment pumping, and chiral hinge
  states in crystalline insulators}}, }\href {\doibase
  10.1103/PhysRevB.96.245115} {\bibfield  {journal} {\bibinfo  {journal} {Phys.
  Rev. B}\ }\textbf {\bibinfo {volume} {96}},\ \bibinfo {pages} {245115}
  (\bibinfo {year} {2017}{\natexlab{b}})}\BibitemShut {NoStop}%
\bibitem [{\citenamefont {Schindler}\ \emph
  {et~al.}(2018{\natexlab{a}})\citenamefont {Schindler}, \citenamefont {Wang},
  \citenamefont {Vergniory}, \citenamefont {Cook}, \citenamefont {Murani},
  \citenamefont {Sengupta}, \citenamefont {Kasumov}, \citenamefont {Deblock},
  \citenamefont {Jeon}, \citenamefont {Drozdov}, \citenamefont {Bouchiat},
  \citenamefont {Gu{\'e}ron}, \citenamefont {Yazdani}, \citenamefont
  {Bernevig},\ and\ \citenamefont {Neupert}}]{Schindler18NP}%
  \BibitemOpen
  \bibfield  {author} {\bibinfo {author} {\bibfnamefont {F.}~\bibnamefont
  {Schindler}}, \bibinfo {author} {\bibfnamefont {Z.}~\bibnamefont {Wang}},
  \bibinfo {author} {\bibfnamefont {M.~G.}\ \bibnamefont {Vergniory}}, \bibinfo
  {author} {\bibfnamefont {A.~M.}\ \bibnamefont {Cook}}, \bibinfo {author}
  {\bibfnamefont {A.}~\bibnamefont {Murani}}, \bibinfo {author} {\bibfnamefont
  {S.}~\bibnamefont {Sengupta}},  \emph {et~al.},\ }\bibfield  {title}
  {\enquote {\bibinfo {title} {Higher-order topology in bismuth}}, }\href
  {\doibase 10.1038/s41567-018-0224-7} {\bibfield  {journal} {\bibinfo
  {journal} {Nat. Phys.}\ }\textbf {\bibinfo {volume} {14}},\ \bibinfo {pages}
  {918} (\bibinfo {year} {2018}{\natexlab{a}})}\BibitemShut {NoStop}%
\bibitem [{\citenamefont {Imhof}\ \emph {et~al.}(2018)\citenamefont {Imhof},
  \citenamefont {Berger}, \citenamefont {Bayer}, \citenamefont {Brehm},
  \citenamefont {Molenkamp}, \citenamefont {Kiessling}, \citenamefont
  {Schindler}, \citenamefont {Lee}, \citenamefont {Greiter}, \citenamefont
  {Neupert},\ and\ \citenamefont {Thomale}}]{Imhof18np}%
  \BibitemOpen
  \bibfield  {author} {\bibinfo {author} {\bibfnamefont {S.}~\bibnamefont
  {Imhof}}, \bibinfo {author} {\bibfnamefont {C.}~\bibnamefont {Berger}},
  \bibinfo {author} {\bibfnamefont {F.}~\bibnamefont {Bayer}}, \bibinfo
  {author} {\bibfnamefont {J.}~\bibnamefont {Brehm}}, \bibinfo {author}
  {\bibfnamefont {L.~W.}\ \bibnamefont {Molenkamp}}, \bibinfo {author}
  {\bibfnamefont {T.}~\bibnamefont {Kiessling}},  \emph {et~al.},\ }\bibfield
  {title} {\enquote {\bibinfo {title} {Topolectrical-circuit realization of
  topological corner modes}}, }\href {\doibase 10.1038/s41567-018-0246-1}
  {\bibfield  {journal} {\bibinfo  {journal} {Nat. Phys.}\ }\textbf {\bibinfo
  {volume} {14}},\ \bibinfo {pages} {925} (\bibinfo {year} {2018})}\BibitemShut
  {NoStop}%
\bibitem [{\citenamefont {Serra-Garcia}\ \emph {et~al.}(2018)\citenamefont
  {Serra-Garcia}, \citenamefont {Peri}, \citenamefont {S{\"u}sstrunk},
  \citenamefont {Bilal}, \citenamefont {Larsen}, \citenamefont {Villanueva},\
  and\ \citenamefont {Huber}}]{Serra-Garcia18nature}%
  \BibitemOpen
  \bibfield  {author} {\bibinfo {author} {\bibfnamefont {M.}~\bibnamefont
  {Serra-Garcia}}, \bibinfo {author} {\bibfnamefont {V.}~\bibnamefont {Peri}},
  \bibinfo {author} {\bibfnamefont {R.}~\bibnamefont {S{\"u}sstrunk}}, \bibinfo
  {author} {\bibfnamefont {O.~R.}\ \bibnamefont {Bilal}}, \bibinfo {author}
  {\bibfnamefont {T.}~\bibnamefont {Larsen}}, \bibinfo {author} {\bibfnamefont
  {L.~G.}\ \bibnamefont {Villanueva}}, \ and\ \bibinfo {author} {\bibfnamefont
  {S.~D.}\ \bibnamefont {Huber}},\ }\bibfield  {title} {\enquote {\bibinfo
  {title} {Observation of a phononic quadrupole topological insulator}}, }\href
  {\doibase 10.1038/nature25156} {\bibfield  {journal} {\bibinfo  {journal}
  {Nature}\ }\textbf {\bibinfo {volume} {555}},\ \bibinfo {pages} {342}
  (\bibinfo {year} {2018})}\BibitemShut {NoStop}%
\bibitem [{\citenamefont {Xie}\ \emph {et~al.}(2019)\citenamefont {Xie},
  \citenamefont {Su}, \citenamefont {Wang}, \citenamefont {Su}, \citenamefont
  {Shen}, \citenamefont {Zhan}, \citenamefont {Lu}, \citenamefont {Wang},\ and\
  \citenamefont {Chen}}]{XieBY19prl}%
  \BibitemOpen
  \bibfield  {author} {\bibinfo {author} {\bibfnamefont {B.-Y.}\ \bibnamefont
  {Xie}}, \bibinfo {author} {\bibfnamefont {G.-X.}\ \bibnamefont {Su}},
  \bibinfo {author} {\bibfnamefont {H.-F.}\ \bibnamefont {Wang}}, \bibinfo
  {author} {\bibfnamefont {H.}~\bibnamefont {Su}}, \bibinfo {author}
  {\bibfnamefont {X.-P.}\ \bibnamefont {Shen}}, \bibinfo {author}
  {\bibfnamefont {P.}~\bibnamefont {Zhan}}, \bibinfo {author} {\bibfnamefont
  {M.-H.}\ \bibnamefont {Lu}}, \bibinfo {author} {\bibfnamefont {Z.-L.}\
  \bibnamefont {Wang}}, \ and\ \bibinfo {author} {\bibfnamefont {Y.-F.}\
  \bibnamefont {Chen}},\ }\bibfield  {title} {\enquote {\bibinfo {title}
  {Visualization of higher-order topological insulating phases in
  two-dimensional dielectric photonic crystals}}, }\href {\doibase
  10.1103/PhysRevLett.122.233903} {\bibfield  {journal} {\bibinfo  {journal}
  {Phys. Rev. Lett.}\ }\textbf {\bibinfo {volume} {122}},\ \bibinfo {pages}
  {233903} (\bibinfo {year} {2019})}\BibitemShut {NoStop}%
\bibitem [{\citenamefont {Qi}\ \emph {et~al.}(2020)\citenamefont {Qi},
  \citenamefont {Qiu}, \citenamefont {Xiao}, \citenamefont {He}, \citenamefont
  {Ke},\ and\ \citenamefont {Liu}}]{QiY20prl}%
  \BibitemOpen
  \bibfield  {author} {\bibinfo {author} {\bibfnamefont {Y.}~\bibnamefont
  {Qi}}, \bibinfo {author} {\bibfnamefont {C.}~\bibnamefont {Qiu}}, \bibinfo
  {author} {\bibfnamefont {M.}~\bibnamefont {Xiao}}, \bibinfo {author}
  {\bibfnamefont {H.}~\bibnamefont {He}}, \bibinfo {author} {\bibfnamefont
  {M.}~\bibnamefont {Ke}}, \ and\ \bibinfo {author} {\bibfnamefont
  {Z.}~\bibnamefont {Liu}},\ }\bibfield  {title} {\enquote {\bibinfo {title}
  {Acoustic realization of quadrupole topological insulators}}, }\href
  {\doibase 10.1103/PhysRevLett.124.206601} {\bibfield  {journal} {\bibinfo
  {journal} {Phys. Rev. Lett.}\ }\textbf {\bibinfo {volume} {124}},\ \bibinfo
  {pages} {206601} (\bibinfo {year} {2020})}\BibitemShut {NoStop}%
\bibitem [{\citenamefont {Ni}\ \emph {et~al.}(2019)\citenamefont {Ni},
  \citenamefont {Weiner}, \citenamefont {Al{\`u}},\ and\ \citenamefont
  {Khanikaev}}]{Ni19nm}%
  \BibitemOpen
  \bibfield  {author} {\bibinfo {author} {\bibfnamefont {X.}~\bibnamefont
  {Ni}}, \bibinfo {author} {\bibfnamefont {M.}~\bibnamefont {Weiner}}, \bibinfo
  {author} {\bibfnamefont {A.}~\bibnamefont {Al{\`u}}}, \ and\ \bibinfo
  {author} {\bibfnamefont {A.~B.}\ \bibnamefont {Khanikaev}},\ }\bibfield
  {title} {\enquote {\bibinfo {title} {Observation of higher-order topological
  acoustic states protected by generalized chiral symmetry}}, }\href {\doibase
  10.1038/s41563-018-0252-9} {\bibfield  {journal} {\bibinfo  {journal} {Nat.
  Mater.}\ }\textbf {\bibinfo {volume} {18}},\ \bibinfo {pages} {113} (\bibinfo
  {year} {2019})}\BibitemShut {NoStop}%
\bibitem [{\citenamefont {Chen}\ \emph {et~al.}(2019)\citenamefont {Chen},
  \citenamefont {Deng}, \citenamefont {Shi}, \citenamefont {Zhao},
  \citenamefont {Chen},\ and\ \citenamefont {Dong}}]{ChenXD19prl}%
  \BibitemOpen
  \bibfield  {author} {\bibinfo {author} {\bibfnamefont {X.-D.}\ \bibnamefont
  {Chen}}, \bibinfo {author} {\bibfnamefont {W.-M.}\ \bibnamefont {Deng}},
  \bibinfo {author} {\bibfnamefont {F.-L.}\ \bibnamefont {Shi}}, \bibinfo
  {author} {\bibfnamefont {F.-L.}\ \bibnamefont {Zhao}}, \bibinfo {author}
  {\bibfnamefont {M.}~\bibnamefont {Chen}}, \ and\ \bibinfo {author}
  {\bibfnamefont {J.-W.}\ \bibnamefont {Dong}},\ }\bibfield  {title} {\enquote
  {\bibinfo {title} {Direct observation of corner states in second-order
  topological photonic crystal slabs}}, }\href {\doibase
  10.1103/PhysRevLett.122.233902} {\bibfield  {journal} {\bibinfo  {journal}
  {Phys. Rev. Lett.}\ }\textbf {\bibinfo {volume} {122}},\ \bibinfo {pages}
  {233902} (\bibinfo {year} {2019})}\BibitemShut {NoStop}%
  \bibitem [{\citenamefont {Peterson}\ \emph {et~al.}(2018)\citenamefont
  {Peterson}, \citenamefont {Benalcazar}, \citenamefont {Hughes},\ and\
  \citenamefont {Bahl}}]{Peterson18nature}%
  \BibitemOpen
  \bibfield  {author} {\bibinfo {author} {\bibfnamefont {C.~W.}\ \bibnamefont
  {Peterson}}, \bibinfo {author} {\bibfnamefont {W.~A.}\ \bibnamefont
  {Benalcazar}}, \bibinfo {author} {\bibfnamefont {T.~L.}\ \bibnamefont
  {Hughes}}, \ and\ \bibinfo {author} {\bibfnamefont {G.}~\bibnamefont
  {Bahl}},\ }\bibfield  {title} {\enquote {\bibinfo {title} {A quantized
  microwave quadrupole insulator with topologically protected corner states}},
  }\href {\doibase 10.1038/nature25777} {\bibfield  {journal} {\bibinfo
  {journal} {Nature}\ }\textbf {\bibinfo {volume} {555}},\ \bibinfo {pages}
  {346} (\bibinfo {year} {2018})}\BibitemShut {NoStop}%
\bibitem [{\citenamefont {Langbehn}\ \emph {et~al.}(2017)\citenamefont
  {Langbehn}, \citenamefont {Peng}, \citenamefont {Trifunovic}, \citenamefont
  {von Oppen},\ and\ \citenamefont {Brouwer}}]{Langbehn17prl}%
  \BibitemOpen
  \bibfield  {author} {\bibinfo {author} {\bibfnamefont {J.}~\bibnamefont
  {Langbehn}}, \bibinfo {author} {\bibfnamefont {Y.}~\bibnamefont {Peng}},
  \bibinfo {author} {\bibfnamefont {L.}~\bibnamefont {Trifunovic}}, \bibinfo
  {author} {\bibfnamefont {F.}~\bibnamefont {von Oppen}}, \ and\ \bibinfo
  {author} {\bibfnamefont {P.~W.}\ \bibnamefont {Brouwer}},\ }\bibfield
  {title} {\enquote {\bibinfo {title} {Reflection-symmetric second-order
  topological insulators and superconductors}}, }\href {\doibase
  10.1103/PhysRevLett.119.246401} {\bibfield  {journal} {\bibinfo  {journal}
  {Phys. Rev. Lett.}\ }\textbf {\bibinfo {volume} {119}},\ \bibinfo {pages}
  {246401} (\bibinfo {year} {2017})}\BibitemShut {NoStop}%
\bibitem [{\citenamefont {Song}\ \emph {et~al.}(2017)\citenamefont {Song},
  \citenamefont {Fang},\ and\ \citenamefont {Fang}}]{SongZD17prl}%
  \BibitemOpen
  \bibfield  {author} {\bibinfo {author} {\bibfnamefont {Z.}~\bibnamefont
  {Song}}, \bibinfo {author} {\bibfnamefont {Z.}~\bibnamefont {Fang}}, \ and\
  \bibinfo {author} {\bibfnamefont {C.}~\bibnamefont {Fang}},\ }\bibfield
  {title} {\enquote {\bibinfo {title} {$(d\ensuremath{-}2)$-dimensional edge
  states of rotation symmetry protected topological states}}, }\href {\doibase
  10.1103/PhysRevLett.119.246402} {\bibfield  {journal} {\bibinfo  {journal}
  {Phys. Rev. Lett.}\ }\textbf {\bibinfo {volume} {119}},\ \bibinfo {pages}
  {246402} (\bibinfo {year} {2017})}\BibitemShut {NoStop}%
\bibitem [{\citenamefont {Schindler}\ \emph
  {et~al.}(2018{\natexlab{b}})\citenamefont {Schindler}, \citenamefont {Cook},
  \citenamefont {Vergniory}, \citenamefont {Wang}, \citenamefont {Parkin},
  \citenamefont {Bernevig},\ and\ \citenamefont {Neupert}}]{Schindler18SA}%
  \BibitemOpen
  \bibfield  {author} {\bibinfo {author} {\bibfnamefont {F.}~\bibnamefont
  {Schindler}}, \bibinfo {author} {\bibfnamefont {A.~M.}\ \bibnamefont {Cook}},
  \bibinfo {author} {\bibfnamefont {M.~G.}\ \bibnamefont {Vergniory}}, \bibinfo
  {author} {\bibfnamefont {Z.}~\bibnamefont {Wang}}, \bibinfo {author}
  {\bibfnamefont {S.~S.~P.}\ \bibnamefont {Parkin}}, \bibinfo {author}
  {\bibfnamefont {B.~A.}\ \bibnamefont {Bernevig}}, \ and\ \bibinfo {author}
  {\bibfnamefont {T.}~\bibnamefont {Neupert}},\ }\bibfield  {title} {\enquote
  {\bibinfo {title} {Higher-order topological insulators}}, }\href
  {https://advances.sciencemag.org/content/4/6/eaat0346} {\bibfield  {journal}
  {\bibinfo  {journal} {Science Advances}\ }\textbf {\bibinfo {volume} {4}}
  (\bibinfo {year} {2018}{\natexlab{b}})}\BibitemShut {NoStop}%
\bibitem [{\citenamefont {Ezawa}(2018)}]{Ezawa18prl}%
  \BibitemOpen
  \bibfield  {author} {\bibinfo {author} {\bibfnamefont {M.}~\bibnamefont
  {Ezawa}},\ }\bibfield  {title} {\enquote {\bibinfo {title} {Higher-order
  topological insulators and semimetals on the breathing kagome and pyrochlore
  lattices}}, }\href {\doibase 10.1103/PhysRevLett.120.026801} {\bibfield
  {journal} {\bibinfo  {journal} {Phys. Rev. Lett.}\ }\textbf {\bibinfo
  {volume} {120}},\ \bibinfo {pages} {026801} (\bibinfo {year}
  {2018})}\BibitemShut {NoStop}%
\bibitem [{\citenamefont {Geier}\ \emph {et~al.}(2018)\citenamefont {Geier},
  \citenamefont {Trifunovic}, \citenamefont {Hoskam},\ and\ \citenamefont
  {Brouwer}}]{Geier18prb}%
  \BibitemOpen
  \bibfield  {author} {\bibinfo {author} {\bibfnamefont {M.}~\bibnamefont
  {Geier}}, \bibinfo {author} {\bibfnamefont {L.}~\bibnamefont {Trifunovic}},
  \bibinfo {author} {\bibfnamefont {M.}~\bibnamefont {Hoskam}}, \ and\ \bibinfo
  {author} {\bibfnamefont {P.~W.}\ \bibnamefont {Brouwer}},\ }\bibfield
  {title} {\enquote {\bibinfo {title} {Second-order topological insulators and
  superconductors with an order-two crystalline symmetry}}, }\href {\doibase
  10.1103/PhysRevB.97.205135} {\bibfield  {journal} {\bibinfo  {journal} {Phys.
  Rev. B}\ }\textbf {\bibinfo {volume} {97}},\ \bibinfo {pages} {205135}
  (\bibinfo {year} {2018})}\BibitemShut {NoStop}%
\bibitem [{\citenamefont {Trifunovic}\ \emph {et~al.}(2019)\citenamefont {Trifunovic}, \citenamefont {Hoskam},\ and\ \citenamefont
  {Brouwer}}]{Trifunovic19prx}%
  \BibitemOpen
  \bibfield  {author} {\bibinfo {author} {\bibfnamefont {L.}~\bibnamefont {Trifunovic}}, \ and\ \bibinfo
  {author} {\bibfnamefont {P.~W.}\ \bibnamefont {Brouwer}},\ }\bibfield
  {title} {\enquote {\bibinfo {title} {Higher-Order Bulk-Boundary Correspondence for Topological Crystalline Phases}}, }\href {\doibase 10.1103/PhysRevX.9.011012} {\bibfield  {journal} {\bibinfo  {journal} {Phys.
  Rev. X}\ }\textbf {\bibinfo {volume} {9}},\ \bibinfo {pages} {011012}
  (\bibinfo {year} {2019})}\BibitemShut {NoStop}%
\bibitem [{\citenamefont {Liu}\ \emph {et~al.}(2019)\citenamefont {Liu},
  \citenamefont {Deng},\ and\ \citenamefont {Wakabayashi}}]{LiuF19prl}%
  \BibitemOpen
  \bibfield  {author} {\bibinfo {author} {\bibfnamefont {F.}~\bibnamefont
  {Liu}}, \bibinfo {author} {\bibfnamefont {H.-Y.}\ \bibnamefont {Deng}}, \
  and\ \bibinfo {author} {\bibfnamefont {K.}~\bibnamefont {Wakabayashi}},\
  }\bibfield  {title} {\enquote {\bibinfo {title} {Helical topological edge
  states in a quadrupole phase}}, }\href {\doibase
  10.1103/PhysRevLett.122.086804} {\bibfield  {journal} {\bibinfo  {journal}
  {Phys. Rev. Lett.}\ }\textbf {\bibinfo {volume} {122}},\ \bibinfo {pages}
  {086804} (\bibinfo {year} {2019})}\BibitemShut {NoStop}%
\bibitem [{\citenamefont {Luo}\ and\ \citenamefont {Zhang}(2019)}]{XWLuo19prl}%
  \BibitemOpen
  \bibfield  {author} {\bibinfo {author} {\bibfnamefont {X.-W.}\ \bibnamefont
  {Luo}}\ and\ \bibinfo {author} {\bibfnamefont {C.}~\bibnamefont {Zhang}},\
  }\bibfield  {title} {\enquote {\bibinfo {title} {Higher-order topological
  corner states induced by gain and loss}}, }\href {\doibase
  10.1103/PhysRevLett.123.073601} {\bibfield  {journal} {\bibinfo  {journal}
  {Phys. Rev. Lett.}\ }\textbf {\bibinfo {volume} {123}},\ \bibinfo {pages}
  {073601} (\bibinfo {year} {2019})}\BibitemShut {NoStop}%
\bibitem [{\citenamefont {Ghorashi}\ \emph {et~al.}(2020)\citenamefont
  {Ghorashi}, \citenamefont {Li},\ and\ \citenamefont
  {Hughes}}]{Ghorashi20prl}%
  \BibitemOpen
  \bibfield  {author} {\bibinfo {author} {\bibfnamefont {S.~A.~A.}\
  \bibnamefont {Ghorashi}}, \bibinfo {author} {\bibfnamefont {T.}~\bibnamefont
  {Li}}, \ and\ \bibinfo {author} {\bibfnamefont {T.~L.}\ \bibnamefont
  {Hughes}},\ }\bibfield  {title} {\enquote {\bibinfo {title} {Higher-order
  Weyl semimetals}}, }\href {\doibase 10.1103/PhysRevLett.125.266804}
  {\bibfield  {journal} {\bibinfo  {journal} {Phys. Rev. Lett.}\ }\textbf
  {\bibinfo {volume} {125}},\ \bibinfo {pages} {266804} (\bibinfo {year}
  {2020})}\BibitemShut {NoStop}%
\bibitem [{\citenamefont {Wang}\ \emph {et~al.}(2020)\citenamefont {Wang},
  \citenamefont {Lin}, \citenamefont {Jiang}, \citenamefont {Guo},\ and\
  \citenamefont {Jiang}}]{WangHX20prl}%
  \BibitemOpen
  \bibfield  {author} {\bibinfo {author} {\bibfnamefont {H.-X.}\ \bibnamefont
  {Wang}}, \bibinfo {author} {\bibfnamefont {Z.-K.}\ \bibnamefont {Lin}},
  \bibinfo {author} {\bibfnamefont {B.}~\bibnamefont {Jiang}}, \bibinfo
  {author} {\bibfnamefont {G.-Y.}\ \bibnamefont {Guo}}, \ and\ \bibinfo
  {author} {\bibfnamefont {J.-H.}\ \bibnamefont {Jiang}},\ }\bibfield  {title}
  {\enquote {\bibinfo {title} {Higher-order Weyl semimetals}}, }\href {\doibase
  10.1103/PhysRevLett.125.146401} {\bibfield  {journal} {\bibinfo  {journal}
  {Phys. Rev. Lett.}\ }\textbf {\bibinfo {volume} {125}},\ \bibinfo {pages}
  {146401} (\bibinfo {year} {2020})}\BibitemShut {NoStop}%
\bibitem [{\citenamefont {Kudo}\ \emph {et~al.}(2019)\citenamefont {Kudo},
  \citenamefont {Yoshida},\ and\ \citenamefont {Hatsugai}}]{Kudo19prl}%
  \BibitemOpen
  \bibfield  {author} {\bibinfo {author} {\bibfnamefont {K.}~\bibnamefont
  {Kudo}}, \bibinfo {author} {\bibfnamefont {T.}~\bibnamefont {Yoshida}}, \
  and\ \bibinfo {author} {\bibfnamefont {Y.}~\bibnamefont {Hatsugai}},\
  }\bibfield  {title} {\enquote {\bibinfo {title} {Higher-order topological
  Mott insulators}}, }\href {\doibase 10.1103/PhysRevLett.123.196402}
  {\bibfield  {journal} {\bibinfo  {journal} {Phys. Rev. Lett.}\ }\textbf
  {\bibinfo {volume} {123}},\ \bibinfo {pages} {196402} (\bibinfo {year}
  {2019})}\BibitemShut {NoStop}%
\bibitem [{\citenamefont {Volpez}\ \emph {et~al.}(2019)\citenamefont {Volpez},
  \citenamefont {Loss},\ and\ \citenamefont {Klinovaja}}]{Volpez19prl}%
  \BibitemOpen
  \bibfield  {author} {\bibinfo {author} {\bibfnamefont {Y.}~\bibnamefont
  {Volpez}}, \bibinfo {author} {\bibfnamefont {D.}~\bibnamefont {Loss}}, \ and\
  \bibinfo {author} {\bibfnamefont {J.}~\bibnamefont {Klinovaja}},\ }\bibfield
  {title} {\enquote {\bibinfo {title} {Second-order topological
  superconductivity in $\ensuremath{\pi}$-junction rashba layers}}, }\href
  {\doibase 10.1103/PhysRevLett.122.126402} {\bibfield  {journal} {\bibinfo
  {journal} {Phys. Rev. Lett.}\ }\textbf {\bibinfo {volume} {122}},\ \bibinfo
  {pages} {126402} (\bibinfo {year} {2019})}\BibitemShut {NoStop}%
\bibitem [{\citenamefont {Wang}\ \emph {et~al.}(2019)\citenamefont {Wang},
  \citenamefont {Wieder}, \citenamefont {Li}, \citenamefont {Yan},\ and\
  \citenamefont {Bernevig}}]{WangZJ19prl}%
  \BibitemOpen
  \bibfield  {author} {\bibinfo {author} {\bibfnamefont {Z.}~\bibnamefont
  {Wang}}, \bibinfo {author} {\bibfnamefont {B.~J.}\ \bibnamefont {Wieder}},
  \bibinfo {author} {\bibfnamefont {J.}~\bibnamefont {Li}}, \bibinfo {author}
  {\bibfnamefont {B.}~\bibnamefont {Yan}}, \ and\ \bibinfo {author}
  {\bibfnamefont {B.~A.}\ \bibnamefont {Bernevig}},\ }\bibfield  {title}
  {\enquote {\bibinfo {title} {Higher-order topology, monopole nodal lines, and
  the origin of large fermi arcs in transition metal dichalcogenides
  $x{\mathrm{te}}_{2}$ ($x=\mathrm{Mo},\mathrm{W}$)}}, }\href {\doibase
  10.1103/PhysRevLett.123.186401} {\bibfield  {journal} {\bibinfo  {journal}
  {Phys. Rev. Lett.}\ }\textbf {\bibinfo {volume} {123}},\ \bibinfo {pages}
  {186401} (\bibinfo {year} {2019})}\BibitemShut {NoStop}%
\bibitem [{\citenamefont {Yan}\ \emph {et~al.}(2018)\citenamefont {Yan},
  \citenamefont {Song},\ and\ \citenamefont {Wang}}]{YanZB18prl}%
  \BibitemOpen
  \bibfield  {author} {\bibinfo {author} {\bibfnamefont {Z.}~\bibnamefont
  {Yan}}, \bibinfo {author} {\bibfnamefont {F.}~\bibnamefont {Song}}, \ and\
  \bibinfo {author} {\bibfnamefont {Z.}~\bibnamefont {Wang}},\ }\bibfield
  {title} {\enquote {\bibinfo {title} {Majorana corner modes in a
  high-temperature platform}}, }\href {\doibase 10.1103/PhysRevLett.121.096803}
  {\bibfield  {journal} {\bibinfo  {journal} {Phys. Rev. Lett.}\ }\textbf
  {\bibinfo {volume} {121}},\ \bibinfo {pages} {096803} (\bibinfo {year}
  {2018})}\BibitemShut {NoStop}%
\bibitem [{\citenamefont {Zhang}\ \emph
  {et~al.}(2020{\natexlab{a}})\citenamefont {Zhang}, \citenamefont {Wu},\ and\
  \citenamefont {Das~Sarma}}]{ZhangRX20prl}%
  \BibitemOpen
  \bibfield  {author} {\bibinfo {author} {\bibfnamefont {R.-X.}\ \bibnamefont
  {Zhang}}, \bibinfo {author} {\bibfnamefont {F.}~\bibnamefont {Wu}}, \ and\
  \bibinfo {author} {\bibfnamefont {S.}~\bibnamefont {Das~Sarma}},\ }\bibfield
  {title} {\enquote {\bibinfo {title} {M\"obius insulator and higher-order
  topology in MnBi$_{2n}$Te$_{3n+1}$}}, }\href {\doibase
  10.1103/PhysRevLett.124.136407} {\bibfield  {journal} {\bibinfo  {journal}
  {Phys. Rev. Lett.}\ }\textbf {\bibinfo {volume} {124}},\ \bibinfo {pages}
  {136407} (\bibinfo {year} {2020}{\natexlab{a}})}\BibitemShut {NoStop}%
\bibitem [{\citenamefont {Li}\ \emph {et~al.}(2020)\citenamefont {Li},
  \citenamefont {Fu}, \citenamefont {Hu}, \citenamefont {Li},\ and\
  \citenamefont {Shen}}]{LiCA20prl}%
  \BibitemOpen
  \bibfield  {author} {\bibinfo {author} {\bibfnamefont {C.-A.}\ \bibnamefont
  {Li}}, \bibinfo {author} {\bibfnamefont {B.}~\bibnamefont {Fu}}, \bibinfo
  {author} {\bibfnamefont {Z.-A.}\ \bibnamefont {Hu}}, \bibinfo {author}
  {\bibfnamefont {J.}~\bibnamefont {Li}}, \ and\ \bibinfo {author}
  {\bibfnamefont {S.-Q.}\ \bibnamefont {Shen}},\ }\bibfield  {title} {\enquote
  {\bibinfo {title} {Topological phase transitions in disordered electric
  quadrupole insulators}}, }\href {\doibase 10.1103/PhysRevLett.125.166801}
  {\bibfield  {journal} {\bibinfo  {journal} {Phys. Rev. Lett.}\ }\textbf
  {\bibinfo {volume} {125}},\ \bibinfo {pages} {166801} (\bibinfo {year}
  {2020})}\BibitemShut {NoStop}%
\bibitem [{\citenamefont {Chen}\ \emph {et~al.}(2020)\citenamefont {Chen},
  \citenamefont {Chen}, \citenamefont {Gao}, \citenamefont {Zhou},\ and\
  \citenamefont {Xu}}]{ChenR20prl}%
  \BibitemOpen
  \bibfield  {author} {\bibinfo {author} {\bibfnamefont {R.}~\bibnamefont
  {Chen}}, \bibinfo {author} {\bibfnamefont {C.-Z.}\ \bibnamefont {Chen}},
  \bibinfo {author} {\bibfnamefont {J.-H.}\ \bibnamefont {Gao}}, \bibinfo
  {author} {\bibfnamefont {B.}~\bibnamefont {Zhou}}, \ and\ \bibinfo {author}
  {\bibfnamefont {D.-H.}\ \bibnamefont {Xu}},\ }\bibfield  {title} {\enquote
  {\bibinfo {title} {Higher-order topological insulators in quasicrystals}},
  }\href {\doibase 10.1103/PhysRevLett.124.036803} {\bibfield  {journal}
  {\bibinfo  {journal} {Phys. Rev. Lett.}\ }\textbf {\bibinfo {volume} {124}},\
  \bibinfo {pages} {036803} (\bibinfo {year} {2020})}\BibitemShut {NoStop}%
\bibitem [{\citenamefont {Yang}\ \emph {et~al.}(2021)\citenamefont {Yang},
  \citenamefont {Li}, \citenamefont {Duan},\ and\ \citenamefont
  {Xu}}]{YangYB21prb}%
  \BibitemOpen
  \bibfield  {author} {\bibinfo {author} {\bibfnamefont {Y.-B.}\ \bibnamefont
  {Yang}}, \bibinfo {author} {\bibfnamefont {K.}~\bibnamefont {Li}}, \bibinfo
  {author} {\bibfnamefont {L.-M.}\ \bibnamefont {Duan}}, \ and\ \bibinfo
  {author} {\bibfnamefont {Y.}~\bibnamefont {Xu}},\ }\bibfield  {title}
  {\enquote {\bibinfo {title} {Higher-order topological Anderson insulators}},
  }\href {\doibase 10.1103/PhysRevB.103.085408} {\bibfield  {journal} {\bibinfo
   {journal} {Phys. Rev. B}\ }\textbf {\bibinfo {volume} {103}},\ \bibinfo
  {pages} {085408} (\bibinfo {year} {2021})}\BibitemShut {NoStop}%
\bibitem [{\citenamefont {Zhang}\ \emph
  {et~al.}(2020{\natexlab{b}})\citenamefont {Zhang}, \citenamefont {Rui},
  \citenamefont {Calzona}, \citenamefont {Choi}, \citenamefont {Schnyder},\
  and\ \citenamefont {Trauzettel}}]{SBZhang20PRR}%
  \BibitemOpen
  \bibfield  {author} {\bibinfo {author} {\bibfnamefont {S.-B.}\ \bibnamefont
  {Zhang}}, \bibinfo {author} {\bibfnamefont {W.~B.}\ \bibnamefont {Rui}},
  \bibinfo {author} {\bibfnamefont {A.}~\bibnamefont {Calzona}}, \bibinfo
  {author} {\bibfnamefont {S.-J.}\ \bibnamefont {Choi}}, \bibinfo {author}
  {\bibfnamefont {A.~P.}\ \bibnamefont {Schnyder}}, \ and\ \bibinfo {author}
  {\bibfnamefont {B.}~\bibnamefont {Trauzettel}},\ }\bibfield  {title}
  {\enquote {\bibinfo {title} {Topological and holonomic quantum computation
  based on second-order topological superconductors}}, }\href {\doibase
  10.1103/PhysRevResearch.2.043025} {\bibfield  {journal} {\bibinfo  {journal}
  {Phys. Rev. Research}\ }\textbf {\bibinfo {volume} {2}},\ \bibinfo {pages}
  {043025} (\bibinfo {year} {2020}{\natexlab{b}})}\BibitemShut {NoStop}%
\bibitem [{\citenamefont {Li}\ \emph
  {et~al.}(2020{\natexlab{b}})\citenamefont {Li}, \citenamefont {Geier},
  \citenamefont {Ingham}, \
  and\ \citenamefont {Scammell}}]{LiT21arxiv}%
  \BibitemOpen
  \bibfield  {author} {\bibinfo {author} {\bibfnamefont {T.}\ \bibnamefont
  {Li}}, \bibinfo {author} {\bibfnamefont {M.}\ \bibnamefont {Geier}},
  \bibinfo {author} {\bibfnamefont {J.}~\bibnamefont {Ingham}},\ and\ \bibinfo {author}
  {\bibfnamefont {H. D.}~\bibnamefont {Scammell}},\ }\bibfield  {title}
  {\enquote {\bibinfo {title} {Higher-order topological superconductivity from repulsive interactions in kagome and honeycomb systems}}, }\href {https://arxiv.org/abs/2108.10897} {\bibfield  {journal} {\bibinfo  {journal}
  {arXiv:}\ }\ \bibinfo {pages}
  {2108.10897} (\bibinfo {year} {2021}{\natexlab{b}})}\BibitemShut {NoStop}%
\bibitem [{\citenamefont {Benalcazar}\ \emph {et~al.}(2019)\citenamefont
  {Benalcazar}, \citenamefont {Li},\ and\ \citenamefont
  {Hughes}}]{Benalcazar19prb}%
  \BibitemOpen
  \bibfield  {author} {\bibinfo {author} {\bibfnamefont {W.~A.}\ \bibnamefont
  {Benalcazar}}, \bibinfo {author} {\bibfnamefont {T.}~\bibnamefont {Li}}, \
  and\ \bibinfo {author} {\bibfnamefont {T.~L.}\ \bibnamefont {Hughes}},\
  }\bibfield  {title} {\enquote {\bibinfo {title} {Quantization of fractional
  corner charge in ${C}_{n}$-symmetric higher-order topological crystalline
  insulators}}, }\href {\doibase 10.1103/PhysRevB.99.245151} {\bibfield
  {journal} {\bibinfo  {journal} {Phys. Rev. B}\ }\textbf {\bibinfo {volume}
  {99}},\ \bibinfo {pages} {245151} (\bibinfo {year} {2019})}\BibitemShut
  {NoStop}%
\bibitem [{\citenamefont {Krutoff}\ \emph {et~al.}(2017)\citenamefont
  {de Boer}, \citenamefont {van Wezel},\citenamefont {Kane},\citenamefont {Kane},\ and\ \citenamefont
  {Hughes}}]{Krutoff17prx}%
  \BibitemOpen
  \bibfield  {author} {\bibinfo {author} {\bibfnamefont {J.}\ \bibnamefont
  {Krutoff}}, \bibinfo {author} {\bibfnamefont {J.}~\bibnamefont {de Boer}},\bibinfo {author} {\bibfnamefont {J.}~\bibnamefont {van Wezel}}, \bibinfo {author} {\bibfnamefont {C. L.}~\bibnamefont {Kane}}, \
  and\ \bibinfo {author} {\bibfnamefont {R.}\ \bibnamefont {Slager}},\
  }\bibfield  {title} {\enquote {\bibinfo {title} {Topological Classification of Crystalline Insulators through Band Structure Combinatorics}}, }\href {\doibase 10.1103/PhysRevX.7.041069} {\bibfield
  {journal} {\bibinfo  {journal} {Phys. Rev. X}\ }\textbf {\bibinfo {volume}
  {7}},\ \bibinfo {pages} {041069} (\bibinfo {year} {2017})}\BibitemShut
  {NoStop}%
  \bibitem [{\citenamefont {Po}\ \emph {et~al.}(2017)\citenamefont
  {Vishwanath}, and\ \citenamefont
  {Watanabe}}]{Po17nc}%
  \BibitemOpen
  \bibfield  {author} {\bibinfo {author} {\bibfnamefont {H. C.}\ \bibnamefont
  {Po}}, \bibinfo {author} {\bibfnamefont {A.}~\bibnamefont {Vishwanath}}, \
  and\ \bibinfo {author} {\bibfnamefont {H.}\ \bibnamefont {Watanabe}},\
  }\bibfield  {title} {\enquote {\bibinfo {title} {Symmetry-based indicators of band topology in the 230 space groups}}, }\href {\doibase 10.1038/s41467-017-00133-2} {\bibfield
  {journal} {\bibinfo  {journal} {Nat. Commun.}\ }\textbf {\bibinfo {volume}
  {8}},\ \bibinfo {pages} {50} (\bibinfo {year} {2017})}\BibitemShut
  {NoStop}%
\bibitem [{Li2()}]{Li2021SM}%
  \BibitemOpen
  \href@noop {} {\bibinfo  {journal} {See Supplemental Material, for more details on the properties
of 2D SSH model and the effective band structure
picture for the metal-band insulator transition, which includes
  Refs. \cite{Benalcazar19prb,Benalcazar20prb,Furusaki99prl}, at
  URL}\ }\BibitemShut {NoStop}%
\bibitem [{\citenamefont {Benalcazar}\ and\ \citenamefont
  {Cerjan}(2020)}]{Benalcazar20prb}%
  \BibitemOpen
\bibfield  {journal} {  }\bibfield  {author} {\bibinfo {author} {\bibfnamefont
  {W.~A.}\ \bibnamefont {Benalcazar}}\ and\ \bibinfo {author} {\bibfnamefont
  {A.}~\bibnamefont {Cerjan}},\ }\bibfield  {title} {\enquote {\bibinfo {title}
  {Bound states in the continuum of higher-order topological insulators}},
  }\href {\doibase 10.1103/PhysRevB.101.161116} {\bibfield  {journal} {\bibinfo
   {journal} {Phys. Rev. B}\ }\textbf {\bibinfo {volume} {101}},\ \bibinfo
  {pages} {161116(R)} (\bibinfo {year} {2020})}\BibitemShut {NoStop}%
\bibitem [{\citenamefont {Cerjan}\ \emph {et~al.}(2020)\citenamefont {Cerjan},
  \citenamefont {J\"urgensen}, \citenamefont {Benalcazar}, \citenamefont
  {Mukherjee},\ and\ \citenamefont {Rechtsman}}]{Cerjan21prl}%
  \BibitemOpen
  \bibfield  {author} {\bibinfo {author} {\bibfnamefont {A.}~\bibnamefont
  {Cerjan}}, \bibinfo {author} {\bibfnamefont {M.}~\bibnamefont {J\"urgensen}},
  \bibinfo {author} {\bibfnamefont {W.~A.}\ \bibnamefont {Benalcazar}},
  \bibinfo {author} {\bibfnamefont {S.}~\bibnamefont {Mukherjee}}, \ and\
  \bibinfo {author} {\bibfnamefont {M.~C.}\ \bibnamefont {Rechtsman}},\
  }\bibfield  {title} {\enquote {\bibinfo {title} {Observation of a
  higher-order topological bound state in the continuum}}, }\href {\doibase
  10.1103/PhysRevLett.125.213901} {\bibfield  {journal} {\bibinfo  {journal}
  {Phys. Rev. Lett.}\ }\textbf {\bibinfo {volume} {125}},\ \bibinfo {pages}
  {213901} (\bibinfo {year} {2020})}\BibitemShut {NoStop}%
\bibitem [{Note1()}]{Note1}%
  \BibitemOpen
  \bibinfo {title} {On an experimental note, the feasibility of manipulating
  gauge fluxes in these systems has recently been demonstrated \cite{Linzk21arxiv}}\BibitemShut
  {NoStop}%
\bibitem [{\citenamefont {Wigner}(1951)}]{Wigner51}%
  \BibitemOpen
  \bibfield  {author} {\bibinfo {author} {\bibfnamefont {E.~P.}\ \bibnamefont
  {Wigner}},\ }\bibfield  {title} {\enquote {\bibinfo {title} {On a class of
  analytic functions from the quantum theory of collisions}}, }\href
  {http://www.jstor.org/stable/1969342} {\bibfield  {journal} {\bibinfo
  {journal} {Annals of Mathematics}\ }\textbf {\bibinfo {volume} {53}},\
  \bibinfo {pages} {36} (\bibinfo {year} {1951})}\BibitemShut {NoStop}%
\bibitem [{\citenamefont {Dyson}(1962)}]{Dyson62jmp}%
  \BibitemOpen
  \bibfield  {author} {\bibinfo {author} {\bibfnamefont {F.~J.}\ \bibnamefont
  {Dyson}},\ }\bibfield  {title} {\enquote {\bibinfo {title} {Statistical
  theory of the energy levels of complex systems. i}}, }\href {\doibase
  10.1063/1.1703773} {\bibfield  {journal} {\bibinfo  {journal} {J. Math.
  Phys.}\ }\textbf {\bibinfo {volume} {3}},\ \bibinfo {pages} {140} (\bibinfo
  {year} {1962})}\BibitemShut {NoStop}%
\bibitem [{\citenamefont {Altland}\ \emph {et~al.}(1997)\citenamefont {Altland},
\ and\ \citenamefont {Zirnbauer}}]{Altland97}%
  \BibitemOpen
  \bibfield  {author} {\bibinfo {author} {\bibfnamefont {A.}~\bibnamefont
  {Altland}}, \ and\
  \bibinfo {author} {\bibfnamefont {M.}~\bibnamefont {Zirnbauer}},\ }\bibfield
  {title} {\enquote {\bibinfo {title} {Nonstandard symmetry classes in mesoscopic normal-superconducting hybrid structures}}, }\href {\doibase
  10.1103/PhysRevB.55.1142} {\bibfield  {journal} {\bibinfo  {journal} {Phys.
  Rev. B}\ }\textbf {\bibinfo {volume} {55}},\ \bibinfo {pages} {1142}
  (\bibinfo {year} {1997})}\BibitemShut {NoStop}%
\bibitem [{\citenamefont {Li}\ \emph {et~al.}(2017)\citenamefont {Li},
  \citenamefont {Li},\ and\ \citenamefont {Das~Sarma}}]{LiX17prb}%
  \BibitemOpen
  \bibfield  {author} {\bibinfo {author} {\bibfnamefont {X.}~\bibnamefont
  {Li}}, \bibinfo {author} {\bibfnamefont {X.}~\bibnamefont {Li}}, \ and\
  \bibinfo {author} {\bibfnamefont {S.}~\bibnamefont {Das~Sarma}},\ }\bibfield
  {title} {\enquote {\bibinfo {title} {Mobility edges in one-dimensional
  bichromatic incommensurate potentials}}, }\href {\doibase
  10.1103/PhysRevB.96.085119} {\bibfield  {journal} {\bibinfo  {journal} {Phys.
  Rev. B}\ }\textbf {\bibinfo {volume} {96}},\ \bibinfo {pages} {085119}
  (\bibinfo {year} {2017})}\BibitemShut {NoStop}%
\bibitem [{\citenamefont {Roy}\ \emph {et~al.}(2021)\citenamefont {Roy},
  \citenamefont {Mishra}, \citenamefont {Tanatar},\ and\ \citenamefont
  {Basu}}]{RoyS21prl}%
  \BibitemOpen
  \bibfield  {author} {\bibinfo {author} {\bibfnamefont {S.}~\bibnamefont
  {Roy}}, \bibinfo {author} {\bibfnamefont {T.}~\bibnamefont {Mishra}},
  \bibinfo {author} {\bibfnamefont {B.}~\bibnamefont {Tanatar}}, \ and\
  \bibinfo {author} {\bibfnamefont {S.}~\bibnamefont {Basu}},\ }\bibfield
  {title} {\enquote {\bibinfo {title} {Reentrant localization transition in a
  quasiperiodic chain}}, }\href {\doibase 10.1103/PhysRevLett.126.106803}
  {\bibfield  {journal} {\bibinfo  {journal} {Phys. Rev. Lett.}\ }\textbf
  {\bibinfo {volume} {126}},\ \bibinfo {pages} {106803} (\bibinfo {year}
  {2021})}\BibitemShut {NoStop}%
  \bibitem [{\citenamefont {Padhan}\ \emph {et~al.}(2022)\citenamefont {Padhan},
  \citenamefont {Giri}, \citenamefont {Modal},\ and\ \citenamefont
  {Mishra}}]{Padhan22prb}%
  \BibitemOpen
  \bibfield  {author} {\bibinfo {author} {\bibfnamefont {A.}~\bibnamefont
  {Padhan}}, \bibinfo {author} {\bibfnamefont {M.}~\bibnamefont {Giri}},
  \bibinfo {author} {\bibfnamefont {S.}~\bibnamefont {Modal}}, \ and\
  \bibinfo {author} {\bibfnamefont {T.}~\bibnamefont {Mishra}},\ }\bibfield
  {title} {\enquote {\bibinfo {title} {Emergence of multiple localization transitions in a one-dimensional quasiperiodic lattice}}, }\href {\doibase 10.1103/PhysRevB.105.L220201}
  {\bibfield  {journal} {\bibinfo  {journal} {Phys. Rev. B}\ }\textbf
  {\bibinfo {volume} {105}},\ \bibinfo {pages} {L220201} (\bibinfo {year}
  {2022})}\BibitemShut {NoStop}%
\bibitem [{\citenamefont {Oganesyan}\ and\ \citenamefont
  {Huse}(2007)}]{Oganesyan07prb}%
  \BibitemOpen
  \bibfield  {author} {\bibinfo {author} {\bibfnamefont {V.}~\bibnamefont
  {Oganesyan}}\ and\ \bibinfo {author} {\bibfnamefont {D.~A.}\ \bibnamefont
  {Huse}},\ }\bibfield  {title} {\enquote {\bibinfo {title} {Localization of
  interacting fermions at high temperature}}, }\href {\doibase
  10.1103/PhysRevB.75.155111} {\bibfield  {journal} {\bibinfo  {journal} {Phys.
  Rev. B}\ }\textbf {\bibinfo {volume} {75}},\ \bibinfo {pages} {155111}
  (\bibinfo {year} {2007})}\BibitemShut {NoStop}%
\bibitem [{\citenamefont {Atas}\ \emph {et~al.}(2013)\citenamefont {Atas},
  \citenamefont {Bogomolny}, \citenamefont {Giraud},\ and\ \citenamefont
  {Roux}}]{Atas13prl}%
  \BibitemOpen
  \bibfield  {author} {\bibinfo {author} {\bibfnamefont {Y.~Y.}\ \bibnamefont
  {Atas}}, \bibinfo {author} {\bibfnamefont {E.}~\bibnamefont {Bogomolny}},
  \bibinfo {author} {\bibfnamefont {O.}~\bibnamefont {Giraud}}, \ and\ \bibinfo
  {author} {\bibfnamefont {G.}~\bibnamefont {Roux}},\ }\bibfield  {title}
  {\enquote {\bibinfo {title} {Distribution of the ratio of consecutive level
  spacings in random matrix ensembles}}, }\href {\doibase
  10.1103/PhysRevLett.110.084101} {\bibfield  {journal} {\bibinfo  {journal}
  {Phys. Rev. Lett.}\ }\textbf {\bibinfo {volume} {110}},\ \bibinfo {pages}
  {084101} (\bibinfo {year} {2013})}\BibitemShut {NoStop}%
  \bibitem [{Not({\natexlab{b}})}]{Note-r-coninutity}%
  \BibitemOpen
  \bibfield  {title} { {\bibinfo {title} { Note that in the thermodynamic
limit, the discrete LSR $r_{n}$ becomes a continuous variable $r\in[0,1]$}}}\href@noop {}\BibitemShut
  {NoStop}%
 \bibitem [{Not({\natexlab{b}})}]{Note2}%
  \BibitemOpen
  \bibfield  {title} { {\bibinfo {title} {At the critical point, the distribution $P_{c}(r)$ is
  different from these two limits (see SM \cite {Li2021SM})}}}\href@noop {}\BibitemShut
  {NoStop}%
\bibitem [{Note3()}]{Note3}%
  \BibitemOpen
  \bibfield  {title} {Here, we consider the case with $|t_{x}|<t$ and $|t_{y}|<t$ for
  concreteness. We provide the calculations for other parameter regimes for
  instance with $|t_{x}|>t$ and $|t_{y}|>t$ and for the random-flux model limit with
  $t_x=t_y=t$ in the SM \cite {Li2021SM}}\href@noop {}\BibitemShut
  {NoStop}%
\bibitem [{\citenamefont {Laumann}\ \emph {et~al.}(2014)\citenamefont
  {Laumann}, \citenamefont {Pal},\ and\ \citenamefont
  {Scardicchio}}]{Laumann14prl}%
  \BibitemOpen
  \bibfield  {author} {\bibinfo {author} {\bibfnamefont {C.~R.}\ \bibnamefont
  {Laumann}}, \bibinfo {author} {\bibfnamefont {A.}~\bibnamefont {Pal}}, \ and\
  \bibinfo {author} {\bibfnamefont {A.}~\bibnamefont {Scardicchio}},\
  }\bibfield  {title} {\enquote {\bibinfo {title} {Many-body mobility edge in a
  mean-field quantum spin glass}}, }\href {\doibase
  10.1103/PhysRevLett.113.200405} {\bibfield  {journal} {\bibinfo  {journal}
  {Phys. Rev. Lett.}\ }\textbf {\bibinfo {volume} {113}},\ \bibinfo {pages}
  {200405} (\bibinfo {year} {2014})}\BibitemShut {NoStop}%
\bibitem [{\citenamefont {Luitz}\ \emph {et~al.}(2015)\citenamefont {Luitz},
  \citenamefont {Laflorencie},\ and\ \citenamefont {Alet}}]{Luitz15prb}%
  \BibitemOpen
  \bibfield  {author} {\bibinfo {author} {\bibfnamefont {D.~J.}\ \bibnamefont
  {Luitz}}, \bibinfo {author} {\bibfnamefont {N.}~\bibnamefont {Laflorencie}},
  \ and\ \bibinfo {author} {\bibfnamefont {F.}~\bibnamefont {Alet}},\
  }\bibfield  {title} {\enquote {\bibinfo {title} {Many-body localization edge
  in the random-field Heisenberg chain}}, }\href {\doibase
  10.1103/PhysRevB.91.081103} {\bibfield  {journal} {\bibinfo  {journal} {Phys.
  Rev. B}\ }\textbf {\bibinfo {volume} {91}},\ \bibinfo {pages} {081103(R)}
  (\bibinfo {year} {2015})}\BibitemShut {NoStop}%
\bibitem [{\citenamefont {Luo}\ \emph {et~al.}(2021)\citenamefont {Luo},
  \citenamefont {Ohtsuki},\ and\ \citenamefont {Shindou}}]{LuoX21prl}%
  \BibitemOpen
  \bibfield  {author} {\bibinfo {author} {\bibfnamefont {X.}~\bibnamefont
  {Luo}}, \bibinfo {author} {\bibfnamefont {T.}~\bibnamefont {Ohtsuki}}, \ and\
  \bibinfo {author} {\bibfnamefont {R.}~\bibnamefont {Shindou}},\ }\bibfield
  {title} {\enquote {\bibinfo {title} {Universality classes of the Anderson
  transitions driven by non-Hermitian disorder}}, }\href {\doibase
  10.1103/PhysRevLett.126.090402} {\bibfield  {journal} {\bibinfo  {journal}
  {Phys. Rev. Lett.}\ }\textbf {\bibinfo {volume} {126}},\ \bibinfo {pages}
  {090402} (\bibinfo {year} {2021})}\BibitemShut {NoStop}%
\bibitem [{\citenamefont {Slevin}\ and\ \citenamefont
  {Ohtsuki}(2014)}]{Slevin14njp}%
  \BibitemOpen
  \bibfield  {author} {\bibinfo {author} {\bibfnamefont {K.}~\bibnamefont
  {Slevin}}\ and\ \bibinfo {author} {\bibfnamefont {T.}~\bibnamefont
  {Ohtsuki}},\ }\bibfield  {title} {\enquote {\bibinfo {title} {Critical
  exponent for the Anderson transition in the three-dimensional orthogonal
  universality class}}, }\href {\doibase 10.1088/1367-2630/16/1/015012}
  {\bibfield  {journal} {\bibinfo  {journal} {New Journal of Physics}\ }\textbf
  {\bibinfo {volume} {16}},\ \bibinfo {pages} {015012} (\bibinfo {year}
  {2014})}\BibitemShut {NoStop}%
\bibitem [{\citenamefont {Slevin}\ and\ \citenamefont
  {Ohtsuki}(2009)}]{Slevin09prb}%
  \BibitemOpen
  \bibfield  {author} {\bibinfo {author} {\bibfnamefont {K.}~\bibnamefont
  {Slevin}}\ and\ \bibinfo {author} {\bibfnamefont {T.}~\bibnamefont
  {Ohtsuki}},\ }\bibfield  {title} {\enquote {\bibinfo {title} {Critical
  exponent for the quantum Hall transition}}, }\href {\doibase
  10.1103/PhysRevB.80.041304} {\bibfield  {journal} {\bibinfo  {journal} {Phys.
  Rev. B}\ }\textbf {\bibinfo {volume} {80}},\ \bibinfo {pages} {041304(R)}
  (\bibinfo {year} {2009})}\BibitemShut {NoStop}%
\bibitem [{exponent()}]{Note_exponent}%
  \BibitemOpen
  \bibfield  {title} { {\bibinfo {title} {According to the general theory of critical phenomena, $\nu$ is universal and determined solely by the universality class and the dimension of the system.}}, }\BibitemShut {NoStop}%
\bibitem [{\citenamefont {Evers}\ and\ \citenamefont
  {Mirlin}(2000)}]{Evers00prl}%
  \BibitemOpen
  \bibfield  {author} {\bibinfo {author} {\bibfnamefont {F.}~\bibnamefont
  {Evers}}\ and\ \bibinfo {author} {\bibfnamefont {A.~D.}\ \bibnamefont
  {Mirlin}},\ }\bibfield  {title} {\enquote {\bibinfo {title} {Fluctuations of
  the inverse participation ratio at the Anderson transition}}, }\href
  {\doibase 10.1103/PhysRevLett.84.3690} {\bibfield  {journal} {\bibinfo
  {journal} {Phys. Rev. Lett.}\ }\textbf {\bibinfo {volume} {84}},\ \bibinfo
  {pages} {3690} (\bibinfo {year} {2000})}\BibitemShut {NoStop}%
\bibitem [{\citenamefont {Mirlin}\ and\ \citenamefont
  {Evers}(2000)}]{Mirlin00prb}%
  \BibitemOpen
  \bibfield  {author} {\bibinfo {author} {\bibfnamefont {A.~D.}\ \bibnamefont
  {Mirlin}}\ and\ \bibinfo {author} {\bibfnamefont {F.}~\bibnamefont {Evers}},\
  }\bibfield  {title} {\enquote {\bibinfo {title} {Multifractality and critical
  fluctuations at the Anderson transition}}, }\href {\doibase
  10.1103/PhysRevB.62.7920} {\bibfield  {journal} {\bibinfo  {journal} {Phys.
  Rev. B}\ }\textbf {\bibinfo {volume} {62}},\ \bibinfo {pages} {7920}
  (\bibinfo {year} {2000})}\BibitemShut {NoStop}%
\bibitem [{\citenamefont {Chalker}\ \emph
  {et~al.}(1996{\natexlab{a}})\citenamefont {Chalker}, \citenamefont
  {Kravtsov},\ and\ \citenamefont {Lerner}}]{Chalker96jetp}%
  \BibitemOpen
  \bibfield  {author} {\bibinfo {author} {\bibfnamefont {J.~T.}\ \bibnamefont
  {Chalker}}, \bibinfo {author} {\bibfnamefont {V.~E.}\ \bibnamefont
  {Kravtsov}}, \ and\ \bibinfo {author} {\bibfnamefont {I.~V.}\ \bibnamefont
  {Lerner}},\ }\bibfield  {title} {\enquote {\bibinfo {title} {Spectral
  rigidity and eigenfunction correlations at the Anderson transition}}, }\href
  {\doibase 10.1134/1.567208} {\bibfield  {journal} {\bibinfo  {journal} {J.
  Exp. Theor. Phys.}\ }\textbf {\bibinfo {volume} {64}},\ \bibinfo {pages}
  {386} (\bibinfo {year} {1996}{\natexlab{a}})}\BibitemShut {NoStop}%
\bibitem [{\citenamefont {Chalker}\ \emph
  {et~al.}(1996{\natexlab{b}})\citenamefont {Chalker}, \citenamefont {Lerner},\
  and\ \citenamefont {Smith}}]{Chalker96prl}%
  \BibitemOpen
  \bibfield  {author} {\bibinfo {author} {\bibfnamefont {J.~T.}\ \bibnamefont
  {Chalker}}, \bibinfo {author} {\bibfnamefont {I.~V.}\ \bibnamefont {Lerner}},
  \ and\ \bibinfo {author} {\bibfnamefont {R.~A.}\ \bibnamefont {Smith}},\
  }\bibfield  {title} {\enquote {\bibinfo {title} {Random walks through the
  ensemble: Linking spectral statistics with wave-function correlations in
  disordered metals}}, }\href {\doibase 10.1103/PhysRevLett.77.554} {\bibfield
  {journal} {\bibinfo  {journal} {Phys. Rev. Lett.}\ }\textbf {\bibinfo
  {volume} {77}},\ \bibinfo {pages} {554} (\bibinfo {year}
  {1996}{\natexlab{b}})}\BibitemShut {NoStop}%
\bibitem [{\citenamefont {Ginibre}(1965)}]{Ginibre00jmp}%
  \BibitemOpen
  \bibfield  {author} {\bibinfo {author} {\bibfnamefont {J.}~\bibnamefont
  {Ginibre}},\ }\bibfield  {title} {\enquote {\bibinfo {title} {Statistical
  ensembles of complex, quaternion, and real matrices}}, }\href {\doibase
  10.1063/1.1704292} {\bibfield  {journal} {\bibinfo  {journal} {J. Math.
  Phys.}\ }\textbf {\bibinfo {volume} {6}},\ \bibinfo {pages} {440} (\bibinfo
  {year} {1965})}\BibitemShut {NoStop}%
\bibitem [{Note4()}]{Note4}%
  \BibitemOpen
  \bibfield  {title} {We provide the calculation and analysis for other values of
  $U$ in the SM \cite {Li2021SM}.}\BibitemShut {Stop}%
\bibitem [{\citenamefont {Kang}\ \emph {et~al.}(2019)\citenamefont {Kang},
  \citenamefont {Shiozaki},\ and\ \citenamefont {Cho}}]{Kang19prb}%
  \BibitemOpen
  \bibfield  {author} {\bibinfo {author} {\bibfnamefont {B.}~\bibnamefont
  {Kang}}, \bibinfo {author} {\bibfnamefont {K.}~\bibnamefont {Shiozaki}}, \
  and\ \bibinfo {author} {\bibfnamefont {G.~Y.}\ \bibnamefont {Cho}},\
  }\bibfield  {title} {\enquote {\bibinfo {title} {Many-body order parameters
  for multipoles in solids}}, }\href {\doibase 10.1103/PhysRevB.100.245134}
  {\bibfield  {journal} {\bibinfo  {journal} {Phys. Rev. B}\ }\textbf {\bibinfo
  {volume} {100}},\ \bibinfo {pages} {245134} (\bibinfo {year}
  {2019})}\BibitemShut {NoStop}%
\bibitem [{\citenamefont {Wheeler}\ \emph {et~al.}(2019)\citenamefont
  {Wheeler}, \citenamefont {Wagner},\ and\ \citenamefont
  {Hughes}}]{Wheeler19prb}%
  \BibitemOpen
  \bibfield  {author} {\bibinfo {author} {\bibfnamefont {W.~A.}\ \bibnamefont
  {Wheeler}}, \bibinfo {author} {\bibfnamefont {L.~K.}\ \bibnamefont {Wagner}},
  \ and\ \bibinfo {author} {\bibfnamefont {T.~L.}\ \bibnamefont {Hughes}},\
  }\bibfield  {title} {\enquote {\bibinfo {title} {Many-body electric multipole
  operators in extended systems}}, }\href {\doibase
  10.1103/PhysRevB.100.245135} {\bibfield  {journal} {\bibinfo  {journal}
  {Phys. Rev. B}\ }\textbf {\bibinfo {volume} {100}},\ \bibinfo {pages}
  {245135} (\bibinfo {year} {2019})}\BibitemShut {NoStop}%
\bibitem [{\citenamefont {Roy}(2019)}]{Roy19prr}%
  \BibitemOpen
  \bibfield  {author} {\bibinfo {author} {\bibfnamefont {B.}~\bibnamefont
  {Roy}},\ }\bibfield  {title} {\enquote {\bibinfo {title} {Antiunitary
  symmetry protected higher-order topological phases}}, }\href {\doibase
  10.1103/PhysRevResearch.1.032048} {\bibfield  {journal} {\bibinfo  {journal}
  {Phys. Rev. Research}\ }\textbf {\bibinfo {volume} {1}},\ \bibinfo {pages}
  {032048(R)} (\bibinfo {year} {2019})}\BibitemShut {NoStop}%
\bibitem [{\citenamefont {Peng}\ \emph {et~al.}(2017)\citenamefont {Peng},
  \citenamefont {Bao},\ and\ \citenamefont {von Oppen}}]{PengY17prb}%
  \BibitemOpen
  \bibfield  {author} {\bibinfo {author} {\bibfnamefont {Y.}~\bibnamefont
  {Peng}}, \bibinfo {author} {\bibfnamefont {Y.}~\bibnamefont {Bao}}, \ and\
  \bibinfo {author} {\bibfnamefont {F.}~\bibnamefont {von Oppen}},\ }\bibfield
  {title} {\enquote {\bibinfo {title} {Boundary green functions of topological
  insulators and superconductors}}, }\href {\doibase
  10.1103/PhysRevB.95.235143} {\bibfield  {journal} {\bibinfo  {journal} {Phys.
  Rev. B}\ }\textbf {\bibinfo {volume} {95}},\ \bibinfo {pages} {235143}
  (\bibinfo {year} {2017})}\BibitemShut {NoStop}%
\bibitem [{Note5()}]{Note5}%
  \BibitemOpen
   \bibfield  {title} {Based on the layered lattice structure, the Green's function
  at the $n$-th layer can be expressed as
  $G_{n}=(E-H_{n}-V_{n-1,n}G_{n-1}V_{n-1,n}^{\dagger })^{-1}$, where $H_{n}$ is
  the Hamiltonian for the $n$-th layer and $V_{n-1,n}$ is the hoping between
  the $(n-1)$-th layer and $n$-th layer. After the iteration process, the edge
  Hamiltonian can be obtained as $H_{\protect \mathrm {edge}}=-G_{\protect
  \mathcal {N}}(E=0)^{-1}$, where $\protect \mathcal {N}$ labels the
  edge layer.}\BibitemShut {Stop}%
\bibitem [{\citenamefont {Resta}(1998)}]{Resta98prl}%
  \BibitemOpen
  \bibfield  {author} {\bibinfo {author} {\bibfnamefont {R.}~\bibnamefont
  {Resta}},\ }\bibfield  {title} {\enquote {\bibinfo {title}
  {Quantum-mechanical position operator in extended systems}}, }\href {\doibase
  10.1103/PhysRevLett.80.1800} {\bibfield  {journal} {\bibinfo  {journal}
  {Phys. Rev. Lett.}\ }\textbf {\bibinfo {volume} {80}},\ \bibinfo {pages}
  {1800} (\bibinfo {year} {1998})}\BibitemShut {NoStop}%
\bibitem [{Note6()}]{Note6}%
  \BibitemOpen
  \bibfield  {title} {Note that translational symmetry in the system can be restored by such an average.
  Periodic boundary conditions are imposed in the calculation.}\BibitemShut
  {Stop}%
\bibitem [{\citenamefont {Otaki}\ and\ \citenamefont
  {Fukui}(2019)}]{Otaki19prb}%
  \BibitemOpen
  \bibfield  {author} {\bibinfo {author} {\bibfnamefont {Y.}~\bibnamefont
  {Otaki}}\ and\ \bibinfo {author} {\bibfnamefont {T.}~\bibnamefont {Fukui}},\
  }\bibfield  {title} {\enquote {\bibinfo {title} {Higher-order topological
  insulators in a magnetic field}}, }\href {\doibase
  10.1103/PhysRevB.100.245108} {\bibfield  {journal} {\bibinfo  {journal}
  {Phys. Rev. B}\ }\textbf {\bibinfo {volume} {100}},\ \bibinfo {pages}
  {245108} (\bibinfo {year} {2019})}\BibitemShut {NoStop}%
\bibitem [{\citenamefont {Zuo}\ \emph {et~al.}(2021)\citenamefont {Zuo},
  \citenamefont {Benalcazar},\ and\ \citenamefont {Liu}}]{zuo21arxiv}%
  \BibitemOpen
  \bibfield  {author} {\bibinfo {author} {\bibfnamefont {Z.-W.}\ \bibnamefont
  {Zuo}}, \bibinfo {author} {\bibfnamefont {W.~A.}\ \bibnamefont {Benalcazar}},
  \ and\ \bibinfo {author} {\bibfnamefont {C.-X.}\ \bibnamefont {Liu}},\
  }\bibfield  {title} {\enquote {\bibinfo {title} {Topological phases of the
  dimerized hofstadter butterfly}}, }\href {\doibase 10.1088/1361-6463/ac12f7} {\bibfield
  {journal} {\bibinfo  {journal} {J. Phys. D: Appl. Phys.}\ }\textbf {\bibinfo {volume}
  {54}},\ \bibinfo {pages} {414004} (\bibinfo {year} {2021})}\BibitemShut
  {NoStop}%
\bibitem [{\citenamefont {Hofstadter}(1976)}]{Hofstadter76prb}%
  \BibitemOpen
  \bibfield  {author} {\bibinfo {author} {\bibfnamefont {D.~R.}\ \bibnamefont
  {Hofstadter}},\ }\bibfield  {title} {\enquote {\bibinfo {title} {Energy
  levels and wave functions of bloch electrons in rational and irrational
  magnetic fields}}, }\href {\doibase 10.1103/PhysRevB.14.2239} {\bibfield
  {journal} {\bibinfo  {journal} {Phys. Rev. B}\ }\textbf {\bibinfo {volume}
  {14}},\ \bibinfo {pages} {2239} (\bibinfo {year} {1976})}\BibitemShut
  {NoStop}%
\bibitem [{\citenamefont {Lin}\ \emph {et~al.}(2021)\citenamefont {Lin},
  \citenamefont {Wu}, \citenamefont {Jiang}, \citenamefont {Liu}, \citenamefont
  {Wu},\ and\ \citenamefont {Jiang}}]{Linzk21arxiv}%
  \BibitemOpen
  \bibfield  {author} {\bibinfo {author} {\bibfnamefont {Z.-K.}\ \bibnamefont
  {Lin}}, \bibinfo {author} {\bibfnamefont {Y.}~\bibnamefont {Wu}}, \bibinfo
  {author} {\bibfnamefont {B.}~\bibnamefont {Jiang}}, \bibinfo {author}
  {\bibfnamefont {Y.}~\bibnamefont {Liu}}, \bibinfo {author} {\bibfnamefont
  {S.}~\bibnamefont {Wu}}, \ and\ \bibinfo {author} {\bibfnamefont {J.-H.}\
  \bibnamefont {Jiang}},\ }\bibfield  {title} {\enquote {\bibinfo {title}
  {Single-plaquette gauge flux as a probe of topological phases on lattices}},
  }\href@noop {} {\  (\bibinfo {year} {2021})},\ \Eprint
  {http://arxiv.org/abs/2105.02070} {arXiv:2105.02070 [cond-mat.mtrl-sci]}
  \BibitemShut {NoStop}%
  \bibitem [{\citenamefont {Li}\ \emph {et~al.}(2021)\citenamefont {Li},
  \citenamefont {Yan}, \citenamefont {Gao}, \ and\ \citenamefont {Hu}}]{LiS21arxiv}%
  \BibitemOpen
  \bibfield  {author} {\bibinfo {author} {\bibfnamefont {S.}\ \bibnamefont
  {Li}}, \bibinfo {author} {\bibfnamefont {X. X}~\bibnamefont {Yan}}, \bibinfo
  {author} {\bibfnamefont {J. H.}~\bibnamefont {Gao}}, \ and\ \bibinfo {author} {\bibfnamefont {Y.}\ \bibnamefont {Hu}},\ }\bibfield  {title} {\enquote {\bibinfo {title}
  {Circuit QED simulator of two-dimensional Su-Schrieffer-Hegger model: magnetic field induced topological phase transition in high-order topological insulators}},
  }\href@noop {} {\  (\bibinfo {year} {2021})},\ \Eprint
  {https://arxiv.org/abs/2109.12919} {arXiv:2109.12919 [cond-mat.mtrl-sci]}
  \BibitemShut {NoStop}%
\bibitem [{\citenamefont {Dong}\ \emph {et~al.}(2021)\citenamefont {Dong},
  \citenamefont {Juri\ifmmode \check{c}\else \v{c}\fi{}i\ifmmode~\acute{c}\else
  \'{c}\fi{}},\ and\ \citenamefont {Roy}}]{Roy21prr}%
  \BibitemOpen
  \bibfield  {author} {\bibinfo {author} {\bibfnamefont {J.}~\bibnamefont
  {Dong}}, \bibinfo {author} {\bibfnamefont {V.}~\bibnamefont {Juri\ifmmode
  \check{c}\else \v{c}\fi{}i\ifmmode~\acute{c}\else \'{c}\fi{}}}, \ and\
  \bibinfo {author} {\bibfnamefont {B.}~\bibnamefont {Roy}},\ }\bibfield
  {title} {\enquote {\bibinfo {title} {Topolectric circuits: Theory and
  construction}}, }\href {\doibase 10.1103/PhysRevResearch.3.023056} {\bibfield
   {journal} {\bibinfo  {journal} {Phys. Rev. Research}\ }\textbf {\bibinfo
  {volume} {3}},\ \bibinfo {pages} {023056} (\bibinfo {year}
  {2021})}\BibitemShut {NoStop}%
  \bibitem [{\citenamefont {Zhang}\ \emph {et~al.}(2021)}]{ZhangW21prl}%
  \BibitemOpen
  \bibfield  {author} {\bibinfo {author} {\bibfnamefont {W.}\ \bibnamefont
  {Zhang}}, \bibinfo {author} {\bibfnamefont {D.}\ \bibnamefont {Zou}},
  \bibinfo {author} {\bibfnamefont {Q.}\ \bibnamefont {Pei}}, \bibinfo
  {author} {\bibfnamefont {W.}~\bibnamefont {He}}, \bibinfo {author}
  {\bibfnamefont {J.}\ \bibnamefont {Bao}}, \bibinfo {author}
  {\bibfnamefont {H.J.}~\bibnamefont {Sun}}, \ and\ \bibinfo {author} {\bibfnamefont {X.}\
  \bibnamefont {Zhang}},\ }\bibfield  {title} {\enquote {\bibinfo {title}
  {Experimental Observation of Higher-Order Topological Anderson Insulators}}, }\href {\doibase
  10.1103/PhysRevLett.126.146802} {\bibfield  {journal} {\bibinfo  {journal}
  {Phys. Rev. Lett.}\ }\textbf {\bibinfo {volume} {126}},\ \bibinfo {pages}
  {146802} (\bibinfo {year} {2021})}\BibitemShut {NoStop}%
\end{thebibliography}

%merlin.mbs apsrev4-1.bst 2010-07-25 4.21a (PWD, AO, DPC) hacked
%Control: key (0)
%Control: author (72) initials jnrlst
%Control: editor formatted (1) identically to author
%Control: production of article title (1) required
%Control: page (0) single
%Control: year (1) truncated
%Control: production of eprint (0) enabled
%

\appendix
\numberwithin{equation}{section}\setcounter{figure}{0}\global\long\def\thefigure{S\arabic{figure}}
\global\long\def\thesection{S\arabic{section}}
\global\long\def\thesubsection{\Alph{subsection}}

\begin{widetext}
\begin{center}
\textbf{\large{}Supplemental material for ``Transition from metal to higher-order topological insulator driven by random flux''}{\large{} }
\par\end{center}{\large \par}
%\tableofcontents

\section{Properties of the 2D SSH model}

In this section, we present the band structure and phase diagram of
the 2D Su-Schrieffer-Heeger (SSH) model in the clean limit. The Hamiltonian
in momentum space is given by
\begin{equation}
H_{0}({\bf k})=(t_{x}+t\cos k_{x})\tau_{1}\sigma_{0}-t\sin k_{x}\tau_{2}\sigma_{3}+(t_{y}+t\cos k_{y})\tau_{1}\sigma_{1}-t\sin k_{y}\tau_{1}\sigma_{2},\label{eq:Hamiltonian2}
\end{equation}
where $\tau$ and $\sigma$ are Pauli matrices for different degrees
of freedom within a unit cell; ${\bf k}=\{k_{x},k_{y}\}$ is the 2D
wavevector; $t$ and $t_{x}$ $(t_{y}$) denotes the two staggered
hopping strengths along $x$ $(y)$ directions. We set the lattice
constant to unity for simplicity. \textcolor{black}{Note that $k_{x}$
and $k_{y}$ are decoupled in Eq.~\eqref{eq:Hamiltonian2}. The total
Hamiltonian can be recast as the sum of two SSH models along $x$
and $y$ directions, respectively, $H_{0}(\mathbf{k})=H_{x}(k_{x})+H_{y}(k_{y})$.}\textcolor{red}{{}
}Accordingly, the four matrix terms in Eq.~\eqref{eq:Hamiltonian2}
can be divided into two groups $G_{1}=\{\tau_{1}\sigma_{0},\tau_{2}\sigma_{3}\}$
and $G_{2}=\{\tau_{1}\sigma_{1},\tau_{1}\sigma_{2}\}$. The matrices
are anticommute within each group but are commute between the groups.
Thus, the four energy bands of Eq.~\eqref{eq:Hamiltonian2} can be
found analytically as
\begin{equation}
E_{\eta}^{\pm}=\pm[\epsilon_{x}(k_{x})+(-1)^{\eta}\epsilon_{y}(k_{y})],\label{eq:independentband}
\end{equation}
where $\epsilon_{\alpha}(k_{\alpha})=\sqrt{t_{\alpha}^{2}+2t_{\alpha}t\cos k_{\alpha}+t^{2}}$,
$\alpha\in\{x,y\}$ and $\eta\in\{1,2\}$.

\subsection{Gapless phase with nodal lines}

Due to chiral symmetry, the two bands touch each other at $E=0$,
when
\begin{equation}
\sqrt{t_{x}^{2}+2t_{x}t\cos k_{x}+t^{2}}=\sqrt{t_{y}^{2}+2t_{y}t\cos k_{y}+t^{2}},
\end{equation}
which gives rise to nodal lines in the $k_{x}\text{-}k_{y}$ plane.
These nodal lines are due to the independency of $H_{x}(k_{x})$ and
$H_{y}(k_{y})$. Thus, they are not protected by any specific symmetries.
The condition for a gapless phase is
\begin{align}
|t_{x}^{2}-t_{y}^{2} & |\leq2|t_{y}t|+2|t_{x}t|,\label{eq:touching-condition}
\end{align}
or more concisely,
\begin{equation}
||t_{x}|-|t_{y}||\leq2|t|.
\end{equation}

\begin{figure}
\centering

\includegraphics[clip,width=0.6\columnwidth]{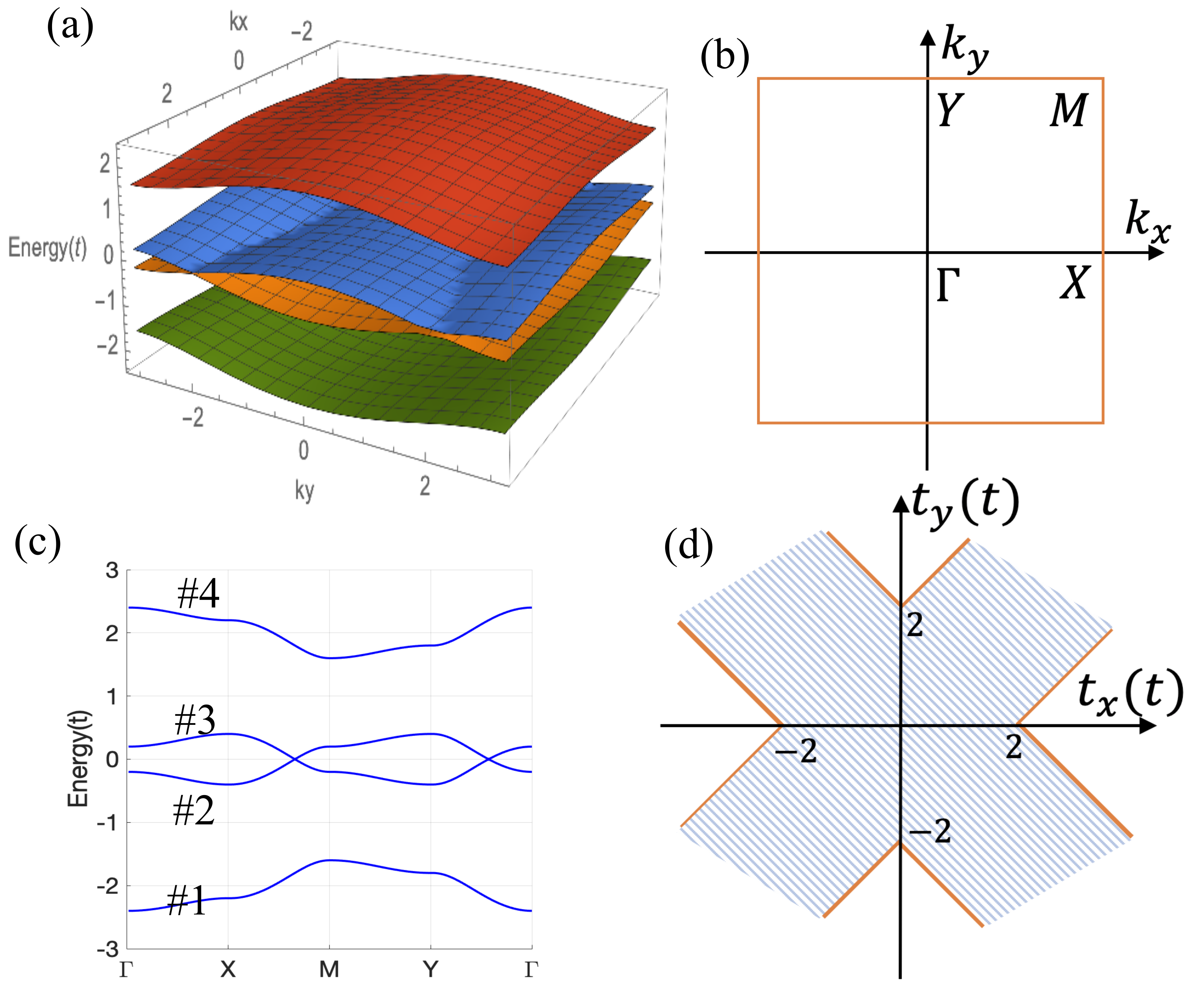}

\caption{(a) Band structures with nodal lines for $t_{x}=0.1t$ and $t_{y}=0.3t$.
(b) Brillouin zone with high symmetry points. (c) Band structure along
the high symmetry lines for $t_{x}=0.1t$ and $t_{y}=0.3t$. (d) Gapless
phase region (shadowed) in the ($t_{x},t_{y}$) parameter space.}

\label{fig:HSP}
\end{figure}

\subsection{Symmetries}

The 2D SSH model possesses a number of symmetries, as listed in Table
\ref{table:symmetry}.

\begin{table}
\centering

\caption{Symmetry properties of the 2D SSH model. Here, $K$ stands for the
complex conjugation.}

\label{table:symmetry}

\begin{tabular}{ccc}
\toprule 
Symmetry & Operator & Operation\tabularnewline
\hline
chiral & $\gamma_5=\tau_{3}\sigma_{0}$ & $\gamma_5 H_{0}({\bf k})\gamma_5^{-1}=-H_{0}({\bf k})$\tabularnewline
\ \ time-reversal\ \  & $\mathcal{T}=\tau_{0}\sigma_{0}K$ & $\mathcal{T}H_{0}({\bf k})\mathcal{T}^{-1}=H_{0}(-{\bf k})$\tabularnewline
particle-hole & $\mathcal{P}=\tau_{3}\sigma_{0}K$ & $\mathcal{P}H_{0}({\bf k})\mathcal{P}^{-1}=-H_{0}(-{\bf k})$\tabularnewline
inversion & $\mathcal{I}=\tau_{0}\sigma_{1}$ & $\mathcal{I}H_{0}({\bf k})\mathcal{I}^{-1}=H_{0}(-{\bf k})$\tabularnewline
mirror $x$ & $M_{x}=\tau_{1}\sigma_{0}$ & \ \ $M_{x}H_{0}(k_{x},k_{y})M_{x}^{-1}=H_{0}(-k_{x},k_{y})$\ \ \tabularnewline
mirror $y$ & $M_{y}=\tau_{1}\sigma_{1}$ & $M_{y}H_{0}(k_{x},k_{y})M_{y}^{-1}=H_{0}(k_{x},-k_{y})$\tabularnewline
PT & $S=\tau_{0}\sigma_{1}K$ & $SH_{0}({\bf k})S^{-1}=H_{0}({\bf k})$\tabularnewline
\hline
\end{tabular}
\end{table}

In general, the 2D SSH model has $C_{2v}$ group symmetry, which includes
mirror symmetries $M_{x}$ and $M_{y}$ with respect to the $x$ and
$y$ axes, respectively, and $C_{2}$ rotation symmetry. The symmetry
representation at the high symmetry points of the Brillouin zone witness
the topological properties (\citep{Benalcazar19prb,Benalcazar20prb}).
The character table of the $C_{2v}$ group is listed in Table \ref{table:C2v}.
There are only one-dimensional irreducible representations. At the
high symmetry points, the wavefunctions form the basis of irreducible
representations of the $C_{2v}$ group, as shown in Table \ref{table:C2v_representations}.
Since the Hamiltonian can be decoupled as \textcolor{black}{$H_{0}(\mathbf{k})=H_{x}(k_{x})+H_{y}(k_{y})$,
we can compare the representations at the $X$ ($Y$) point relative
to the $\Gamma$ point to determine whether the system is topological
or not along the $x$ ($y$) direction. For example, in the case 3
of }Table \ref{table:C2v_representations}, all the bands at the $X$
point have different representations relative to the $\Gamma$ point.
Namely, the representation changes between $A_{(1,2)}$ $\leftrightarrow B_{(1,2)}$
when going from $X$ to $\Gamma$ points, implying a parity shift.
Thus, the system is nontrivial along the $x$ direction. On the other
hand, all the bands at the $Y$ point have the same representations
compared to the $\Gamma$ point. Therefore, the system is trivial
along the $y$ direction. Based on this method, we conclude that the
system is nontrivial along $x$ ($y$) direction when $|t_{x}|<1$
($|t_{y}|<1$). Otherwise, it is trivial. This classification
is consistent with the spectrum of a ribbon along $x$ ($y$) direction,
as shown in Fig. \ref{fig:ribbons band}.

\begin{table}
\centering

\caption{Character table of the $C_{2v}$ group}

\label{table:C2v}

\begin{tabular}{ccccc}
\toprule 
 & \quad{}I\quad{} & \quad{}$C_{2}$\quad{} & \quad{}$M_{x}$\quad{} & \quad{}$M_{y}$\quad{}\tabularnewline
 \hline
$A_{1}$ & 1 & 1 & 1 & 1\tabularnewline
$A_{2}$ & 1 & 1 & -1 & -1\tabularnewline
$B_{1}$ & 1 & -1 & 1 & -1\tabularnewline
$B_{2}$ & 1 & -1 & -1 & 1\tabularnewline
 \hline
\end{tabular}
\end{table}
\begin{table}
\centering

\caption{Symmetry representations of the $C_{2v}$ group at the high symmetry
points in the Brillouin zone. Irreducible representations $A_{1},A_{2},B_{1}$
and $B_{2}$ are all one dimensional.}

\label{table:C2v_representations}

\begin{tabular}{ccccc}
\toprule 
 & \quad{}bands\quad{} & \quad{}$\Gamma$\quad{} & \quad{}$X$\quad{} & \quad{}$Y$\quad{}\tabularnewline
 \hline
case 1: $|t_{x}|<1,|t_{y}|<1$ & \#1 & $A_{2}$ & $B_{1}$ & $B_{2}$\tabularnewline 
 & \#2 & $B_{1}$ & $A_{2}$ & $A_{1}$\tabularnewline
 & \#3 & $B_{2}$ & $A_{1}$ & $A_{2}$\tabularnewline
 & \#4 & $A_{1}$ & $B_{2}$ & $B_{1}$\tabularnewline
 \hline
case 2: $|t_{x}|>1,|t_{y}|>1$ & \#1 & $A_{2}$ & $A_{2}$ & $A_{2}$\tabularnewline
 & \#2 & $B_{1}$ & $B_{1}$ & $B_{1}$\tabularnewline
 & \#3 & $B_{2}$ & $B_{2}$ & $B_{2}$\tabularnewline
 & \#4 & $A_{1}$ & $A_{1}$ & $A_{1}$\tabularnewline
 \hline
case 3: $|t_{x}|<1,|t_{y}|>1$ & \#1 & $A_{2}$ & $B_{1}$ & $A_{2}$\tabularnewline 
 & \#2 & $B_{1}$ & $A_{2}$ & $B_{1}$\tabularnewline
 & \#3 & $B_{2}$ & $A_{1}$ & $B_{2}$\tabularnewline
 & \#4 & $A_{1}$ & $B_{2}$ & $A_{1}$\tabularnewline
 \hline
case 4: $|t_{x}|>1,|t_{y}|<1$ & \#1 & $A_{2}$ & $A_{2}$ & $B_{2}$\tabularnewline
 & \#2 & $B_{2}$ & $B_{2}$ & $A_{2}$\tabularnewline
 & \#3 & $B_{1}$ & $B_{1}$ & $A_{1}$\tabularnewline
 & \#4 & $A_{1}$ & $A_{1}$ & $B_{1}$\tabularnewline
 \hline
\end{tabular}
\end{table}

In the $C_{2v}$ group, there is no 2D irreducible representation.
Consequently, the zero-energy modes stemming from nontrivial topology
(for $|t_{x,y}|<1$) easily hybridize with degenerate bulk states.
Thus, they are unstable. For the special case $t_{x}=t_{y}$, the
system has $C_{4v}$ group symmetry. Because of the 2D irreducible
representation of the $C_{4v}$ group together with chiral symmetry,
the zero-energy modes are protected from the hybridization with the
degenerate bulk states in this case \citep{Benalcazar20prb}.

\begin{figure}
\centering

\includegraphics[clip,width=0.68\columnwidth]{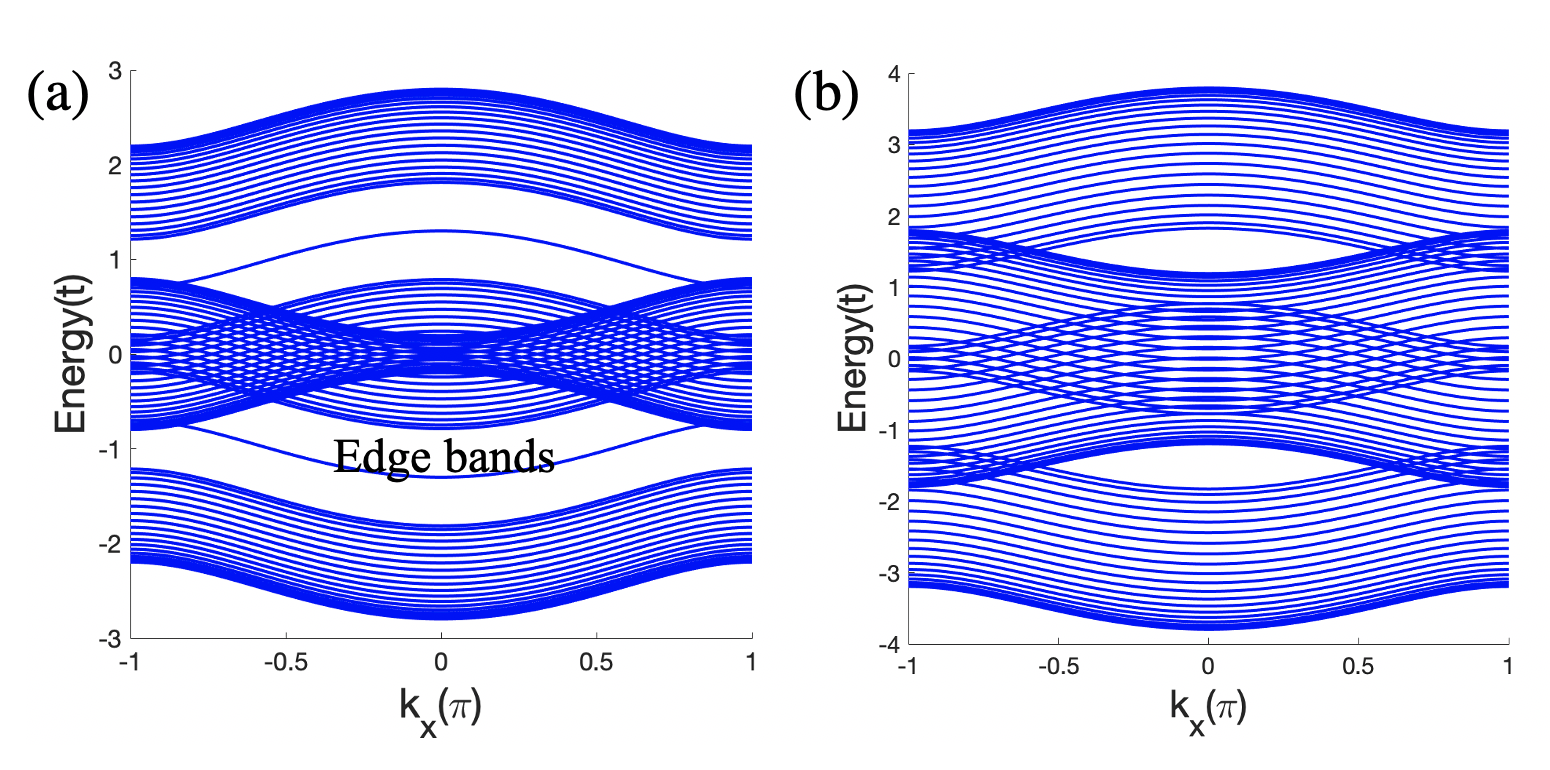}

\caption{(a) Band structure in the $k_{x}$ direction of the model with open
boundaries along the $y$ direction for $t_{x}=0.3t$ and $t_{y}=0.5t$.
The system is nontrivial with edge bands (isolated lines) along the
$y$ direction. (b) Band structure in the $k_{x}$ direction of the
model with open boundaries along the $y$ direction for $t_{x}=0.3t$
and $t_{y}=1.5t$. The system is trivial without edge bands along
$y$ direction.}

\label{fig:ribbons band}
\end{figure}

\section{Metal-insulator transition in the trivial regime}

In this section, we show that the metal-insulator transition by random
flux also occurs in the topologically trivial regime (with $|t_{x(y)}|>t$)
of the 2D SSH model. The results are displayed in Fig.\ \ref{fig:trivial-1}.
Clearly, the random flux opens a band gap beyond a critical random
flux strength $U$, similar to the topological regime. However, there
are no zero-energy corner modes in the insulating phase. Correspondingly
in Fig.\ \ref{fig:trivial-1}(b), the LSR is close to the value $0.6$
for small $U$ and approaches $0.386$ for large $U$. These results
confirm the metal-insulator transition driven by the random flux in
the topologically trivial regime.

\begin{figure}[h]
\centering

\includegraphics[clip,width=0.68\columnwidth]{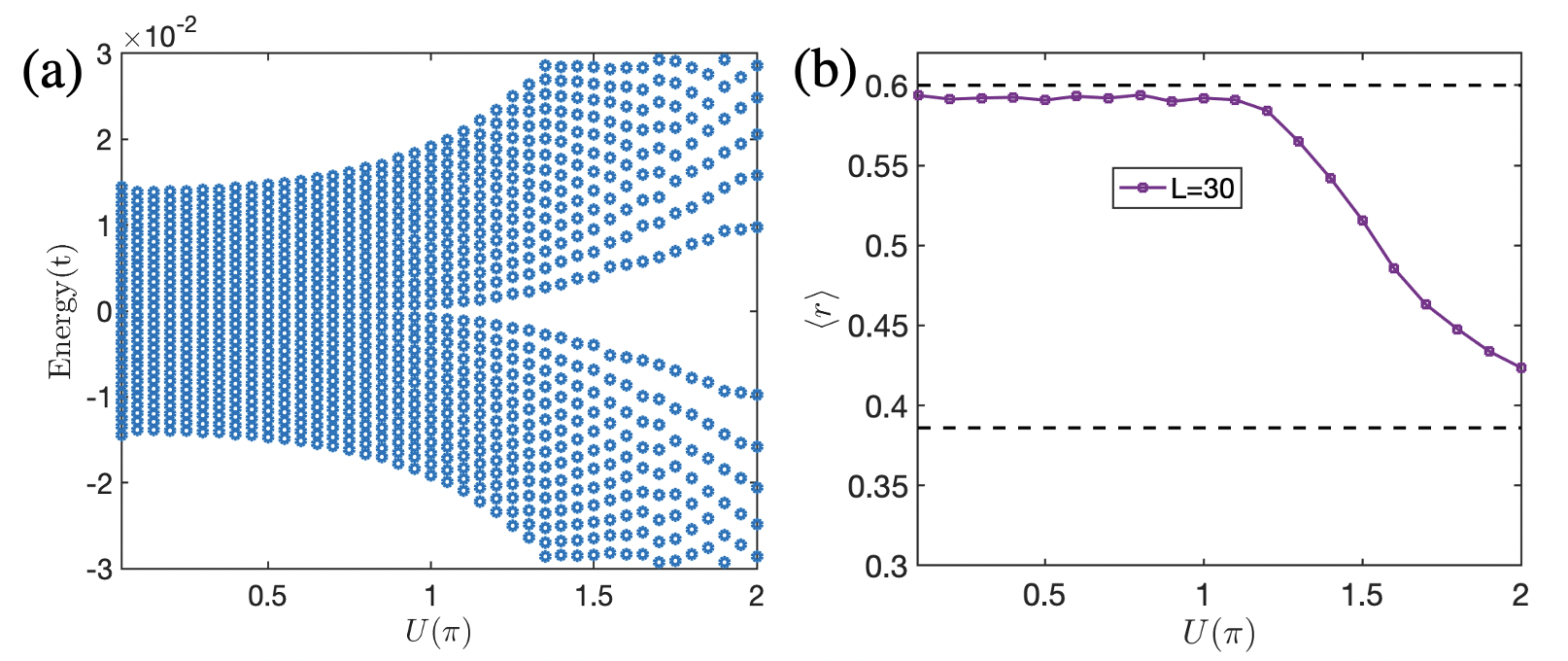}

\caption{(a) Energy spectrum as a function of $U$ for $t_{x}=2t$ and $t_{y}=1.6t$.
There are no degenerate zero-energy states in the insulating phase.
40 energy levels around zero energy are shown. (b) LSR as a function
of $U$ near the band center for $t_{x}=2t$ and $t_{y}=1.6t$. We
take $L=40$ ($30$) and $200(4000)$ disorder configurations in (a){[}(b){]}.
$\langle r\rangle$ at $U=2\pi$ is still away from $0.386$ because
the critical point $U_{c}$ is large and a finite size effect is significant
for $L=40$ (30).}

\label{fig:trivial-1}
\end{figure}

\section{Effective Hamiltonian and band structure}

In this section, we derive and analyze the effective Hamiltonian $H_{\text{eff}}$
and energy spectrum of the system averaged over random flux configurations.
For each random flux configuration, we can find the retarded Green's
function $G^{R}({\bf r},{\bf r}';\omega)$ in the original basis of
tight-binding model as
\begin{alignat}{1}
G^{R}({\bf r},{\bf r}';\omega) & =\langle{\bf r}|(\omega+i\eta-H_{\text{rand}})^{-1}|{\bf r}'\rangle,
\end{alignat}
where $\omega$ is frequency, $\eta$ is an infinitesimal positive
number, and $H_{\text{rand}}$ is the lattice Hamiltonian with the
random flux. We assume periodic boundary conditions in both $x$ and
$y$ directions. Averaging over many random-flux configurations, translation
invariance is effectively restored in both directions. Thus, the disorder-average
Green's function
\begin{alignat}{1}
G_{\mathrm{avg}}^{R}({\bf r}-{\bf r}',\omega) & =\langle G^{R}({\bf r},{\bf r}';\omega)\rangle_{\mathrm{dis}},
\end{alignat}
depends only on ${\bf r}-{\bf r}'$. We perform Fourier transformation
of the disorder-averaged Green's function
\begin{alignat}{1}
G^{R}({\bf k},\omega) & =\int d({\bf r}-{\bf r}')G_{\mathrm{avg}}^{R}({\bf r}-{\bf r}',\omega)e^{i{\bf k}\cdot({\bf r}-{\bf r}')}.\label{eq:RetardedG}
\end{alignat}
Without loss of generality, we replace ${\bf r}-{\bf r}'$ by ${\bf r}$
and consider the discretization of ${\bf r}$ in the lattice model.
Then, Eq.\ \eqref{eq:RetardedG} can be recast as
\begin{alignat}{1}
G^{R}({\bf k},\omega) & =\sum_{{\bf r}}G_{\mathrm{avg}}^{R}({\bf r},\omega)e^{i{\bf k}\cdot{\bf r}}.
\end{alignat}
This procedure allows us to derive the effective Hamiltonian as
\begin{equation}
H_{\mathrm{eff}}({\bf k})=-[G^{R}({\bf k},\omega=0)]^{-1}.
\end{equation}

The corresponding energy band structures for different representative
random-flux strengths are presented in Fig.\ \ref{fig:avgband}.
First, to verify the validity of the disorder-averaged Green's function
method, we show that the results obtained for $U=0$ perfectly agree
with the original energy spectrum {[}see Fig.\ \ref{fig:avgband}(a){]}.
The two middle bands cross each other along nodal lines at zero energy.
Before the metal-insulator transition, the energy spectrum remains
gapless {[}see Fig.\ \ref{fig:avgband}(b), the two middle bands
cross each other at eight points in the Brillouin zone{]}. This corresponds
to small random flux strengths $U<U_{c}$. Note that the critical
strength is $U_{c}\approx0.75\pi$ for the parameters we choose in
the calculation. At the critical point, each two gapless points meet
at the same position in the spectrum {[}see Fig.\ \ref{fig:avgband}(c){]}.
Increasing further random flux strength $U$, the system transmits
from a gapless phase to a gapped phase. This corresponds to the metal-insulator
transition. For $U>U_{c}$, the system becomes fully gapped {[}see
Fig.\ \ref{fig:avgband}(d){]}. Note that the energy may have an
imaginary part (corresponding to finite lifetimes of states) due to
the effective scattering by random disorder. However, this imaginary
part is not important for our analysis.

\begin{figure}[h]
\centering

\includegraphics[clip,width=0.66\columnwidth]{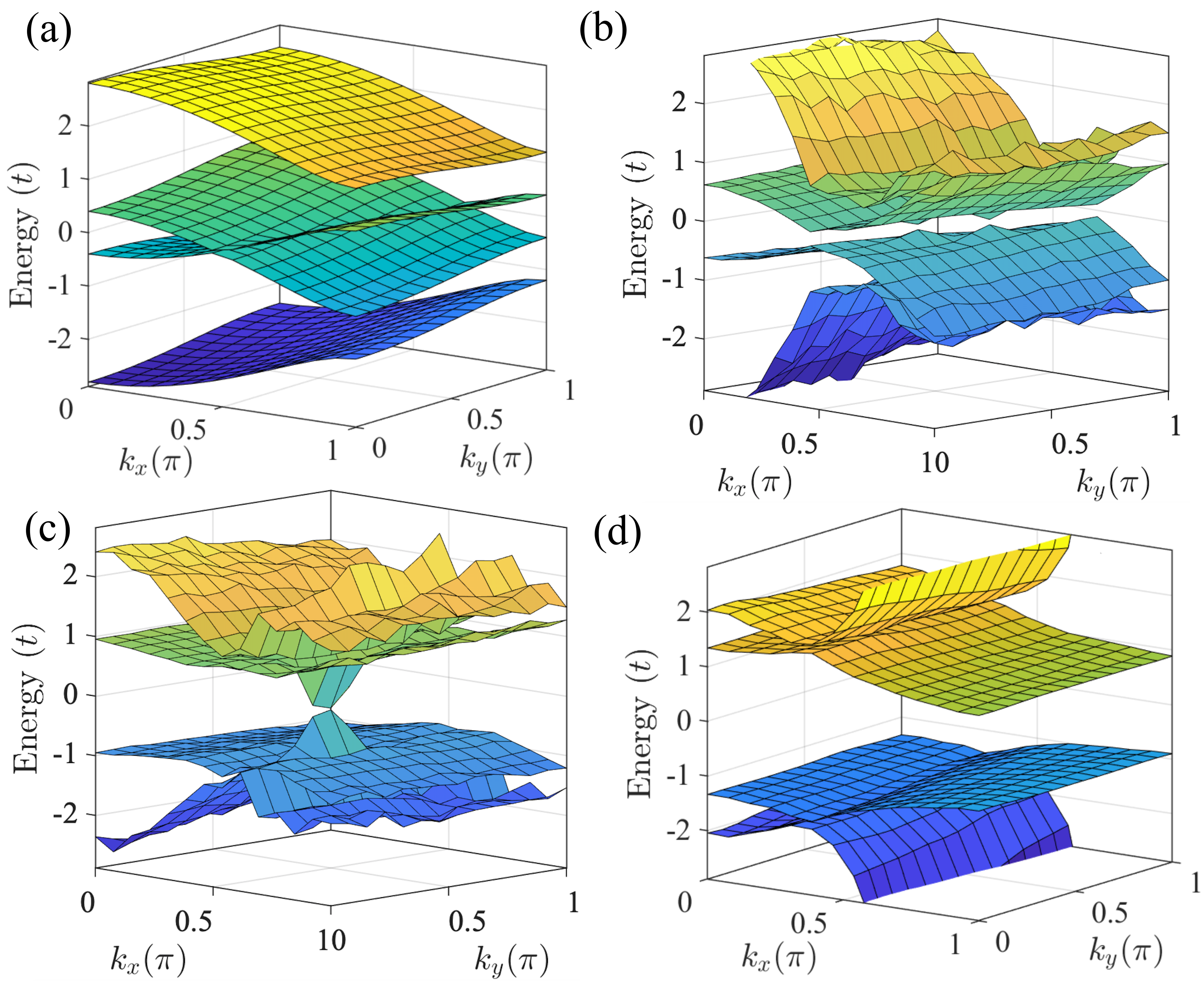}

\caption{Effective band structures obtained by the disorder-averaged Green's
function method for four different random-flux strengths: (a) $U=0$,
(b) $0.3\text{\ensuremath{\pi}}$, (c) $U_{c}=0.75\pi$, and (d) $1.2\pi$,
respectively. $3.8\times10^{5}$ configurations are considered. Other
parameters are $L=26$, $t_{x}=0.2t$ and $t_{y}=0.6t$.}

\label{fig:avgband}
\end{figure}

As shown by our numerical calculations (see Fig.\ 3 in the main text),
the 2D SSH model with random flux can resemble the BBH model. The
ground state of the half-filled 2D SSH model with random-flux strength
$U=2\pi$ (zero mean flux on each plaquette) can be adiabatically
mapped to that of the BBH model (with $\pi$ flux on each plaquette).
To better understand how the random flux influences the system, we
compare the 2D SSH model with the BBH model. In the clean case, the
two models are given respectively by
\begin{alignat}{1}
H_{\mathrm{0}}({\bf k}) & =(t_{x}+t\cos k_{x})\tau_{1}\sigma_{0}-t\sin k_{x}\tau_{2}\sigma_{3}+(t_{y}+t\cos k_{y})\tau_{1}\sigma_{1}-t\sin k_{y}\tau_{1}\sigma_{2},\\
H_{\mathrm{BBH}}({\bf k}) & =(t_{x}+t\cos k_{x})\tau_{1}\sigma_{3}+t\sin k_{x}\tau_{2}\sigma_{0}+(t_{y}+t\cos k_{y})\tau_{1}\sigma_{1}+t\sin k_{y}\tau_{1}\sigma_{2}.
\end{alignat}
The two models share common matrices $\tau_{1}\sigma_{1}$ and $\tau_{1}\sigma_{2}$
(acting on the sublattice degrees of freedom). The 2D SSH model has
two more matrices $\tau_{1}\sigma_{0}$ and $\tau_{2}\sigma_{3}$.
Since the four matrices of the 2D SSH model do not all anticommute,
the bulk system can have a gapless band structure. In contrast, the
BBH model possesses two other additional matrices $\tau_{1}\sigma_{3}$
and $\tau_{2}\sigma_{0}$. Since the four matrices of the BBH model
all anticommute, the bulk system has a fully gapped band structure.
Thus, there are totally six different matrices in the two models.

The resemblance of our model with the BBH model implies that the self-energy
$\Sigma({\bf k})$ due to the presence of random flux may be expanded
in terms of
\begin{equation}
\Sigma({\bf k})=m_{x}\tau_{1}\sigma_{0}+m_{y}\tau_{1}\sigma_{1}+m_{p}\tau_{1}\sigma_{3}+f_{x}\tau_{2}\sigma_{3}+f_{y}\tau_{1}\sigma_{2}+f_{p}\tau_{2}\sigma_{0}.\label{eq: selfenergy}
\end{equation}
We choose this ansatz in the following. Thus, the effective Hamiltonian
of our disorder-averaged system can be written as
\begin{alignat}{1}
H_{\mathrm{eff}}({\bf k}) & =H_{0}({\bf k})+\Sigma({\bf k}).
\end{alignat}
We demonstrate below that self-energy not only modifies the coefficients
of the original matrices in the 2D SSH model but also introduces additional
terms associated with new matrices ($\tau_{1}\sigma_{3}$ and $\tau_{2}\sigma_{0}$)
to the Hamiltonian $H_{0}({\bf k})$. Importantly, it is momentum
dependent and not diagonal in sublattice space. Note that $\Sigma({\bf k})$
may contain non-Hermitian components which give rise to the finite
lifetimes of states. However, they are not important to our results
and thus ignored.

We confirm the above ansatz by calculating $H_{\mathrm{eff}}({\bf k})$
numerically and extracting the corresponding coefficients of $\Sigma({\bf k})$,
as shown in Fig.\ \ref{fig:coefficinets mf}. Clearly, $\Sigma({\bf k})$
effectively modifies the coefficients of the four original matrices
in $H_{0}({\bf k})$ and gives rise to the new terms associated with
$\tau_{1}\sigma_{3}$ and $\tau_{2}\sigma_{0}$. Moreover, the coefficients
in Eq.\ \eqref{eq: selfenergy} are highly momentum-dependent. This
result indicates that the effective Hamiltonian $H_{\mathrm{eff}}({\bf k})$
can be regarded as a mixture of the 2D SSH model and the BBH model.

\begin{figure}[h]
\centering

\includegraphics[clip,width=0.68\columnwidth]{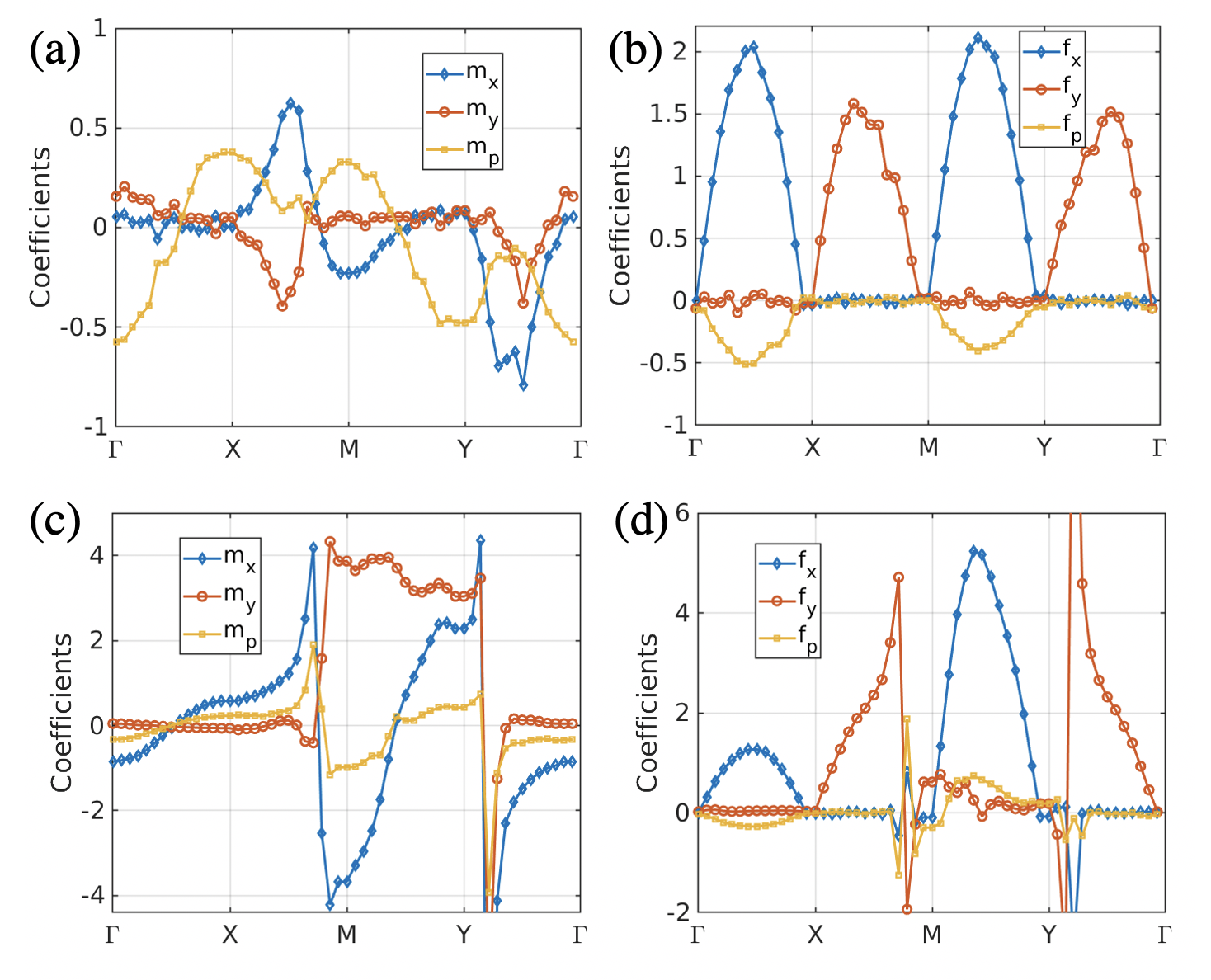}

\caption{(a) Coefficients $m_{x,y,p}$ along the high symmetry lines for $U=0.3\pi$.
(b) Coefficients $f_{x,y,p}$ along the high symmetry lines for $U=0.3\pi$.
(c) Similar to (a) but for $U=1.2\pi$. (d) Similar to (b) but for
$U=1.2\pi$. We choose $L=26$ and $3.8\times10^{5}$ disorder configurations.}

\label{fig:coefficinets mf}
\end{figure}

The modification to the original matrices and the emergence of the
new matrices $\tau_{1}\sigma_{3}$ and $\tau_{2}\sigma_{0}$ in Eq.~\eqref{eq: selfenergy}
can be explained by the scattering of electrons by the random flux.
To do so, we consider, for instance, the contribution (to the self-energy)
from scattering processes of the lowest nonzero order. This contribution
can be calculated as
\begin{equation}
\Sigma_{(2)}({\bf k})=\langle VG_{0}V\rangle_{\text{ave}},\label{eq:self-energy}
\end{equation}
where $G_{0}$ is the bare Green's function in the absence of flux
and given by
\begin{equation}
G_{0}=\dfrac{f_{1}\Gamma_{1}+f_{2}\Gamma_{2}+f_{3}\Gamma_{3}+f_{4}\Gamma_{4}}{f_{1}^{2}+f_{2}^{2}+f_{3}^{2}+f_{4}^{2}},\label{eq:G-function}
\end{equation}
and $V$ is the modification due to the presence of random flux
\begin{equation}
V\equiv H_{\text{rand}}({\bf k},{\bf A})-H_{\mathrm{0}}({\bf k})=t(g_{3}\Gamma_{3}+g_{4}\Gamma_{4}).\label{eq:Potential}
\end{equation}
In the latter equation, ${\bf A}=(A_{x},A_{y})$ is the vector potential
associated with the random flux. For concreteness, we choose the Landau
gauge with $A_{x}=0$ and consider $\omega=0$. $\langle...\rangle_{\text{ave}}$
indicates the average over random flux configurations and the integral
over internal variables; ${\bf k}$ acts as operator $-i\partial_{{\bf r}}$.
We use the following shorthand notations
\begin{align}
\Gamma_{1} & \equiv\tau_{1}\sigma_{0},\ \ \Gamma_{2}=\tau_{2}\sigma_{3},\ \ \Gamma_{3}\equiv\tau_{1}\sigma_{1},\ \ \Gamma_{4}\equiv\tau_{1}\sigma_{2},\nonumber \\
f_{1} & \equiv-(t_{x}+t\cos k_{x}),\ \ f_{2}\equiv t\sin k_{x},\ \ f_{3}\equiv t_{y}+t\cos k_{y},\ \ f_{4}\equiv-t\sin k_{y},\nonumber \\
g_{3} & \equiv\cos(k_{y}-A_{y})-\cos k_{y},\ \ g_{4}\equiv-\sin(k_{y}-A_{y})+\sin k_{y}.
\end{align}
This implies that 
\begin{align}
H_{0}({\bf k}) & =-f_{1}\Gamma_{1}-f_{2}\Gamma_{2}+f_{3}\Gamma_{3}+f_{4}\Gamma_{4},\nonumber \\
V({\bf k},{\bf A}) & =t(g_{3}\Gamma_{3}+g_{4}\Gamma_{4}).
\end{align}
 Plugging Eqs.\ \eqref{eq:G-function} and \eqref{eq:Potential}
into Eq.\ \eqref{eq:self-energy}, we find that
\begin{align}
\Sigma_{(2)}({\bf k}) & =t^{2}\langle(g_{3}\Gamma_{3}+g_{4}\Gamma_{4})(f_{1}\Gamma_{1}+f_{2}\Gamma_{2}+f_{3}\Gamma_{3}+f_{4}\Gamma_{4})\varLambda(g_{3}\Gamma_{3}+g_{4}\Gamma_{4})\rangle_{\text{ave}}\nonumber \\
 & =\langle\mathcal{F}_{1}\rangle\Gamma_{1}+\langle\mathcal{F}_{2}\rangle\Gamma_{2}+\langle\mathcal{F}_{3}\rangle\Gamma_{3}+\langle\mathcal{F}_{4}\rangle\Gamma_{4}+\langle\mathcal{G}_{1}\rangle\tau_{1}\sigma_{3}+\langle\mathcal{G}_{2}\rangle\tau_{2}\sigma_{0},\label{eq:self-energy-result}
\end{align}
where $\varLambda=(f_{1}^{2}+f_{2}^{2}+f_{3}^{2}+f_{4}^{2})^{-1}$
and the corresponding coefficient functions are given by
\begin{align}
\langle\mathcal{F}_{1}\rangle & =t^{2}\langle g_{3}f_{1}\varLambda g_{3}+g_{4}f_{1}\varLambda g_{4}\rangle_{\text{ave}},\nonumber \\
\langle\mathcal{F}_{2}\rangle & =t^{2}\langle g_{3}f_{2}\varLambda g_{3}+g_{4}f_{2}\varLambda g_{4}\rangle_{\text{ave}},\nonumber \\
\langle\mathcal{F}_{3}\rangle & =t^{2}\langle g_{3}f_{3}\varLambda g_{3}+g_{3}f_{4}\varLambda g_{4}+g_{4}f_{4}\varLambda g_{3}-g_{4}f_{3}\varLambda g_{4}\rangle_{\text{ave}},\nonumber \\
\langle\mathcal{F}_{4}\rangle & =t^{2}\langle-g_{3}f_{4}\varLambda g_{3}+g_{3}f_{3}\varLambda g_{4}+g_{4}f_{3}\varLambda g_{3}+g_{4}f_{4}\varLambda g_{4}\rangle_{\text{ave}},\nonumber \\
\langle\mathcal{G}_{1}\rangle & =it^{2}\langle g_{3}f_{1}\varLambda g_{4}-g_{4}f_{1}\varLambda g_{3}\rangle_{\text{ave}},\nonumber \\
\langle\mathcal{G}_{2}\rangle & =it^{2}\langle g_{3}f_{2}\varLambda g_{4}-g_{4}f_{2}\varLambda g_{3}\rangle_{\text{ave}}.
\end{align}
To derive Eq.\ \eqref{eq:self-energy-result}, we employ
\begin{align}
\Gamma_{3}\Gamma_{1}\Gamma_{4} & =-\Gamma_{4}\Gamma_{1}\Gamma_{3}=i\tau_{1}\sigma_{3},\nonumber \\
\Gamma_{3}\Gamma_{2}\Gamma_{4} & =-\Gamma_{4}\Gamma_{2}\Gamma_{3}=i\tau_{2}\sigma_{0}.
\end{align}
Remarkably, $\Sigma_{(2)}({\bf k})$ has the same matrix structure
as the ansatz, Eq.\ \eqref{eq:self-energy}. In a similar way, we
can derive the contributions of higher order scattering processes.
The emergence of the new matrices (i.e., nonzero $\langle\mathcal{G}_{1}\rangle$
and $\langle\mathcal{G}_{2}\rangle$) stems from the interplay between
the intrinsic sublattice degrees of freedom and the fact that the
magnetic flux couples to momentum.

\section{Conventional random flux model}

In this section, we connect our results to the conventional random
flux model which corresponds to $t_{x}=t_{y}=t$ in our model. For
this limit, the model looses its inner degrees of freedom. Consequently,
the spectrum reduces to $E({\bf k})=t(\cos k_{x}+\cos k_{y}),$ which
is the spectrum of a conventional 2D electron gas.

It is known that the band center is delocalized in the conventional
random flux model \citep{Furusaki99prl}. In Fig.\ \ref{fig:2DEG}(a),
we show clearly that the random flux cannot open a bulk gap around
the band center at $E=0$. Correspondingly in Fig.\ \ref{fig:2DEG}(b),
the LSR for states near the band center stays around $0.6$, confirming
its delocalized nature. However, the states away from the band center
are localized.

\begin{figure}[th]
\centering

\includegraphics[clip,width=0.6\columnwidth]{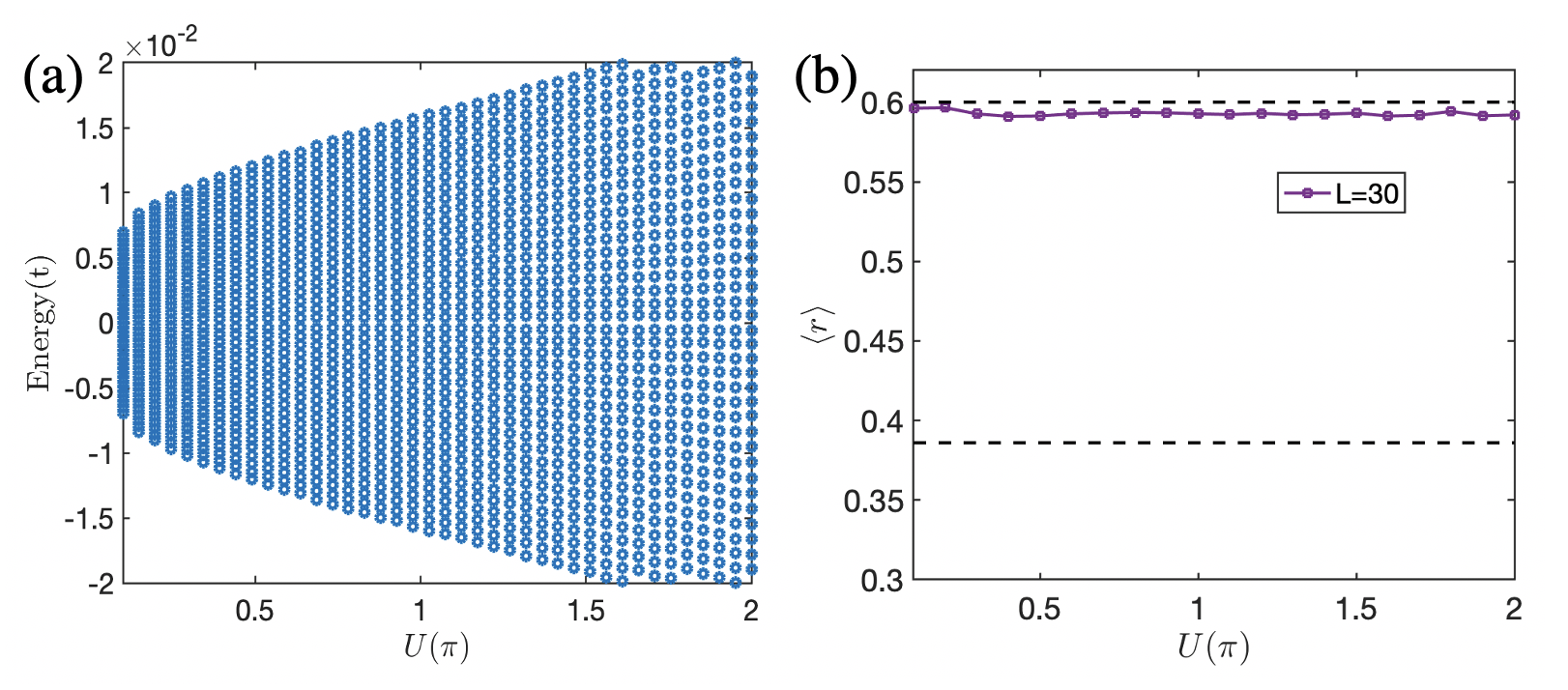}

\caption{(a) Energy spectrum as a function of $U$ for the $t_{x}=t_{y}=t$
limit. 40 energy levels around zero energy are shown. (b) LSR as a
function of $U$ near the band center for the $t_{x}=t_{y}=t$ limit
. We take $L=40$ ($L=30$) and 200 (4000) disorder configurations
in (a) {[}(b){]}.}

\label{fig:2DEG}
\end{figure}

\end{widetext}

\end{document}